\documentclass[useAMS,usenatbib]{mn2e}

\usepackage{amsmath,amssymb}
\usepackage{amsfonts}
\usepackage{bbold}
\usepackage{graphicx}
\usepackage{color}
\usepackage[draft]{hyperref}
\usepackage{rotating}
\usepackage{caption}
\usepackage{natbib}
\usepackage{color}

\usepackage{xspace}

\def\aap{{A\&A}}
\def\mnras{{MNRAS}}
\def\apj{{ApJ}}
\def\apjl{{ApJL}}

\newcommand{\refe}{}

\title[RWI around spinning black holes]{On the Rossby Wave Instability in accretion discs surrounding spinning black holes}
\author[F. Casse \& P. Varniere ]{F. Casse$^{1,2}$\thanks{\href{mailto:fcasse@apc.univ-paris7.fr}{fcasse@apc.univ-paris7.fr}} \& P. Varniere$^{1,2}$ \\
$^{1}$ Laboratoire AstroParticule \& Cosmologie, Sorbonne Paris Cit\'e, Universit\'e Paris Diderot, CNRS/IN2P3, CEA/Irfu, Observatoire\\ de Paris, 10, rue Alice Domon et L\'eonie Duquet, F-75205 Paris Cedex 13, France\\
$^{2}$ AIM, CEA, CNRS, UniversitŽ Paris-Saclay, Universit\'e Paris Diderot, Sorbonne Paris Cit\', F-91191 Gif-sur-Yvette, France.} 

\begin{document}

\date{Accepted YEAR MONTH DAY. Received YEAR MONTH DAY; in original form YEAR MONTH DAY}

\pagerange{\pageref{firstpage}--\pageref{lastpage}} \pubyear{YEAR}

\maketitle

%%\label{firstpage}

\begin{abstract}
We have performed general relativistic hydrodynamics (GRHD) simulations of 2D discs orbiting around spinning black holes and prone to the Rossby Wave Instability (RWI). We  show that the RWI can develop at any location in the disc and for any spin parameter. After recovering the overall patterns of the instability in this general relativistic context, we have analysed its development and identified some modifications induced by the combined effects of the relativistic rotational profile of the disc and local time dilatation  that affects the propagation of waves in the disc. We have found in particular that the saturation level of the instability increases significantly when RWI occurs in the very close vicinity of fast-rotating black holes where general relativistic effects are strong.
  Such finding suggests that even more strongly than in the case of Schwarzschild black-hole, it is necessary to complement such GRHD simulations with a full GR ray-tracing processing in order to  provide synthetic observations of the disc in the distant observer frame. 
\end{abstract}

%%%%%%%%%%%%%%%%%%%%%%%%%%%%%%%%%
\section{Introduction}
The Rossby Wave Instability (RWI) was first used in an astrophysical context by \citet{Lovelace78} for the study of galactic discs and was at the time dubbed as the negative mass instability. 
{Then it was applied to Keplerian disc with analytical studies by \citet{Lovelace99} and \citet{Li00} to derive the stability criterion of this instability.} 
 Following these studies, numerical computations by \citet{Li01} have studied the non-linear stage of the instability and its ability to produce vortices inside the discs as well as accretion of gas.
 Over the last decade, the RWI has received a growing attention from the astrophysical community,  in particular regarding planet-formation \citep{VT06,Meheut10,Lyra12,Lin12,Meheut12}, supermassive black holes \citep{Tagger06,Vin14} and accretion disc orbiting around compact objects  {in particular to explain the origin of fast variability detected in X-ray binary system} \citep{TV06,VTR11,VTR12,Vin13}. 
 {Here we will focus on the later two cases when strong gravity can occur, for a more complete review on recent developments we refer the reader to \citet{Lovelace14}. }

{The first step was presented in} \citet{Cass17} {which} has recently {performed} the first general relativistic hydrodynamics (GRHD) simulations depicting the development of the RWI  within accretion discs orbiting around Schwarzschild black holes. This work has proven the existence of such instability in a full general relativity context and also studied its properties in the gravity field of the non-spinning black hole.  The main result of {that} previous study is that non-relativistic fluid  simulations  taking into account a pseudo-newtonian gravity field gives an adequate description of the instability occurring in discs orbiting around Schwarzschild black holes. In this study we aim at extending our previous computations to the case of Kerr black holes. In such framework we will be able to investigate a much wider variety of configurations in terms of spin parameter and innermost stable orbits radius.
 {Indeed this is of} interest to the study of {the fast variability of} X-ray binaries systems as some of them exhibiting high frequency {Quasi-Periodic Oscillations}  seems to harbour fast rotating black holes \citep{Remillard06}. It is then important to test if previous results regarding RWI around Schwarzschild black holes are still valid when considering fast spinning black holes.  
 
The article is organized as follows: Sect. 2 is devoted to the presentation of the GRHD framework  and simulations set-up. In Sect. 3 we investigate the influence of the spin of the black hole upon the development of the RWI at a fixed distance from the compact object. In Sect. 4 we pursue our analysis of the RWI by allowing the RWI to be triggered closer and closer to the compact object as the spin of the black hole increases. We then compare the differences in the  behaviour of the instability with respect to the location of the RWI. 
 Finally we deliver our conclusions and perspectives in Sect. 5. 
%%%%%%%%%%%%%%%%%%%%%%%%%%%%%%%%%%%%%
%%%%%%%%%%%%%%%%%%%%%%%%%%%%%

\section{Rossby Wave Instability in GRHD framework}
 Assessing the impact of the spin of the black hole requires to perform general relativistic hydrodynamical simulations. We introduce in
 the following subsections the GRHD framework as well as the numerical setup used to achieve such computations using our general relativistic 
 version of  {\tt MPI-AMRVAC} \cite[for more details see][]{Cass17}.

\subsection{GRHD set of equations}

The local geometry of spacetime in the close vicinity of a spinning black hole is described by the Kerr metric \citep{Kerr63}. In such context the general expression of a line element is, using a (3+1) splitting of spacetime in Boyer-Lindquist coordinates $(r,\theta,\varphi)$
\begin{equation} 
\refe{ds^2= -\alpha^2(cdt)^2 + \gamma_{ij}(dx^i+\beta^icdt)(dx^j+\beta^jcdt)\nonumber}
\end{equation}
where $\gamma_{ij}$ is the spatial metric tensor while $\alpha$ is the lapse function and $\beta^i$ is the shift vector \refe{($c$ is the speed of light in vacuum)}. Let us note that latin indices stand for spatial coordinates while greek indices are linked to all spacetime coordinates. The Kerr solution to Einstein equations provides the expression of all previous elements, namely
\begin{equation} 
\refe{\alpha^2=\frac{\xi\Delta}{\Theta}}; \ \gamma_{rr}=\frac{\xi}{\Delta}; \gamma_{\theta\theta}=\xi ; \ \gamma_{\varphi\varphi}=\tilde{\omega}^2
\end{equation}
where the auxiliary functions are $\xi=r^2+a_*^2\cos^2\theta$, $\Delta=r^2-2r_gr+a_*^2$,\ \refe{$\tilde{\omega}^2=\Theta\sin^2\theta/\xi$ and $\Theta=(r^2+a_*^2)^2-a_*^2\Delta\sin^2\theta$}. The angular momentum $J$ of the black hole is directly connected to $a_*=J/Mc=ar_g$ where $M$ is the mass of the black hole and $c$ is the speed of light. We denote $r_g$ as the gravitational radius of the compact object as $r_g=GM/c^2$ where $G$ is the gravitational constant. Let us note that the spin parameter $a$ ranges from $-1$ to $1$. The shift vector in the Kerr solution reads 
\begin{equation}
\beta^i=\left(0,0,\refe{-\frac{2rr_ga_*}{\Theta}}\right)
\end{equation}
which obviously depends on the sign of the spin of the black hole. The gravitational field generated by a spinning black hole depends on  its mass and its spinning parameter $a_*$. The resulting space-time geometry allows the gas of the accretion disc to experience stable circular trajectories closer to the central object than for the Schwarzschild black hole case. Indeed one can easily show that the radius of innermost stable circular orbit (ISCO) in the equatorial plane is provided by the following relations (see e.g. \citet{MTW73})  
\begin{equation}
\frac{r_{\rm ISCO}}{r_g} = 3 + Z_2 \pm\sqrt{(3-Z_1)(3+Z_1+2Z_2)}
\end{equation}
where  $Z_1$ and $Z_2$ are functions depending on $a$, namely
\begin{eqnarray}
Z_1 & = & 1 + (1-a^2)^{1/3}\left((1+a)^{1/3}+(1-a)^{1/3}\right)\nonumber \\
Z_2 & = & \sqrt{3a^2+Z_1^2}
\end{eqnarray}
The $\pm$ sign stands for prograde (smallest $r_{\rm ISCO}$) or retrograde (largest $r_{\rm ISCO}$) orbits. Regarding prograde accretion discs, the  $r_{\rm ISCO}$ ranges from $6 r_g$ for Schwarzschild black holes to $r_g$ for black holes having the maximal spin ($a=1$). 

The set of equations of general relativistic hydrodynamics translates the local conservation of particle number density and relativistic momentum  as follows
 \begin{eqnarray}
 \partial_t (\sqrt{\gamma}D) +\partial_j\left(\sqrt{\gamma}D\left(\alpha\text{v}^j-\beta^j\right)\right) = &0&\nonumber\\
 \partial_t (\sqrt{\gamma}S_i) + \partial_j\left(\sqrt{\gamma}\left[S_i(\alpha\text{v}^j-\beta^j)+\alpha P\delta_i^j\right]\right) = &&\nonumber\\
 \sqrt{\gamma}\left\{-(W^2\rho hc^2-P)\partial_i\alpha + \frac{\alpha}{2}\left(S^j\text{v}^k+P\gamma^{jk}\right)\partial_i\gamma_{jk} \right. && \nonumber\\
 \left. + S_j\partial_i\beta^j \right\}  && 
 \label{SetGRHD}
 \end{eqnarray} 
 where $D=W\rho$ stands as the density of the plasma, $W$ as the Lorentz factor and $\rho$ as the proper density of the fluid.  The covariant velocity of the fluid  is denoted as ${\rm v}^i$ and is related to the Lorentz factor $W=(1-{\rm v}_i{\rm v}^i/c^2)^{-1/2}$. The relativistic momentum is then $S_i=W^2\rho hc {\rm v}_i$ where $\rho hc^2$ is the enthalpy of the fluid. 
  Following \citet{Cass17} we choose to replace the energy conservation equation by a simple \refe{power-law}  linking the thermal pressure $P$ to proper gas density, 
 namely $P=C_o\rho^{\tilde{\gamma}}$ where $C_o$ and $\tilde{\gamma}$ are two positive constants. 
   Relativistic gas thermodynamics prevents the use of a standard polytropic equation of state linking the internal energy of the gas to its density. Following \citet{Meliani04} and \citet{Mignone07}, 
   we can derive such relation by considering the properties of the distribution function of a relativistic gas \citep{Taub48,Mat71}. This leads to  an expression of the internal energy density $u$ such as
  \begin{equation}
  u = \frac{P}{\Gamma -1}+\sqrt{\displaystyle\frac{P^2}{(\Gamma -1)^2}+\rho^2c^4} \nonumber
  \label{Eq:internalU}
   \end{equation}   
  where $\Gamma=5/3$. Then enthalpy can be directly computed from the previous equation since 
  \begin{equation}
  \rho hc^2 = \frac{1}{2}\left\{(\Gamma+1)u-(\Gamma-1)\frac{\rho^2c^4}{u}\right\}\nonumber
  \label{Eq:Enthalpy} 
   \end{equation}  
\refe{Such equation of state is able to accurately describe the transition from low to high temperature gas state where the corresponding polytropic index $\Gamma_{\rm eq}$ ranges between $5/3$ and $4/3$.
It is noteworthy that $\Gamma_\mathrm{eq} \equiv P/(u-\rho c^2)+1$ obtained with this
equation of state only differs by a few percents from that of the theoretical Synge equation \citep{Syn57,Mat71}. As we only consider  thin accretion discs  in this study where $P\ll \rho c^2$ all simulations exhibit  $\Gamma_{\rm eq}$ close to $5/3$.} 
%%%%%%%%%%%%%%%%%%%%%%%%%%%%%%%%%%%%%%%%%%

\subsection{Numerical setup and boundaries}

%%%%%%%%%%%%%%%
The simulations performed in this paper aim at describing the growth of the RWI in an hydrodynamical disc orbiting around a spinning black hole. 
\refe{The thickness of the disc is expected to be very small as in X-ray binaries so we will focus our numerical investigations on vertically integrated discs. In previous studies performed in a Newtonian context, full 3D simulations of the RWI have shown that no vertical mode associated with the RWI has been identified provided that no other instability is at work altogether with RWI \citep{Meheut10,Meheut12}. In \citet{Cass17} we have shown that  for the case $a=0$, going to 3D did not change any of the behaviour of RWI compared to 2D simulations.}  \refe{As the vertical gravity of the Kerr metric is similar for all spin in very thin discs, including the already tested case $a=0$, we can  assume that the 3D behaviour of the RWI}  will stay the same \refe{for higher spins}. This means that the local development of the RWI will mainly depend on the radial structure of the disc. Therefore we restrict ourselves to 2D simulations of the aforementioned discs in the equatorial plane, namely in the $(r,\varphi)$ plane with $\theta=\pi/2$.  
 
\begin{figure}
\includegraphics[width=0.5\textwidth]{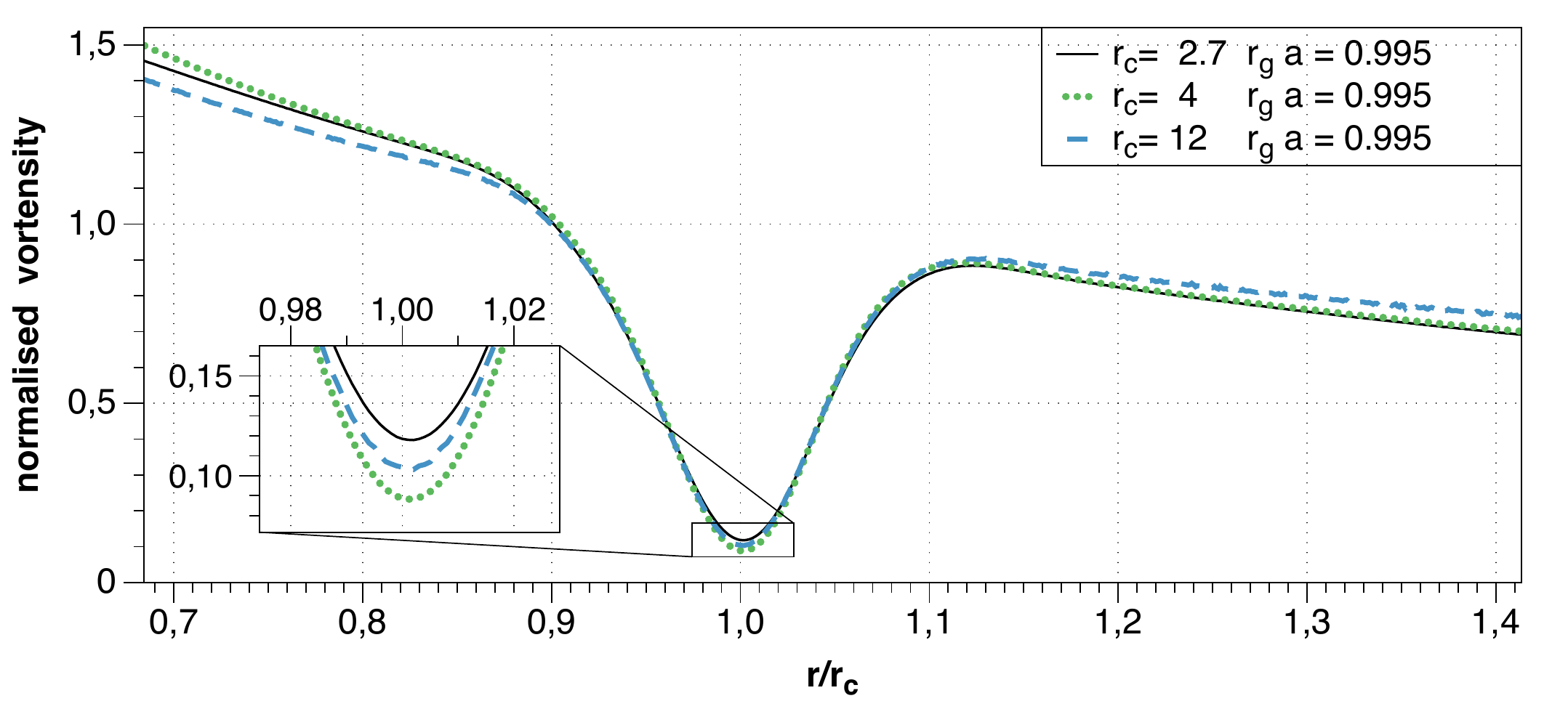}
\includegraphics[width=0.5\textwidth]{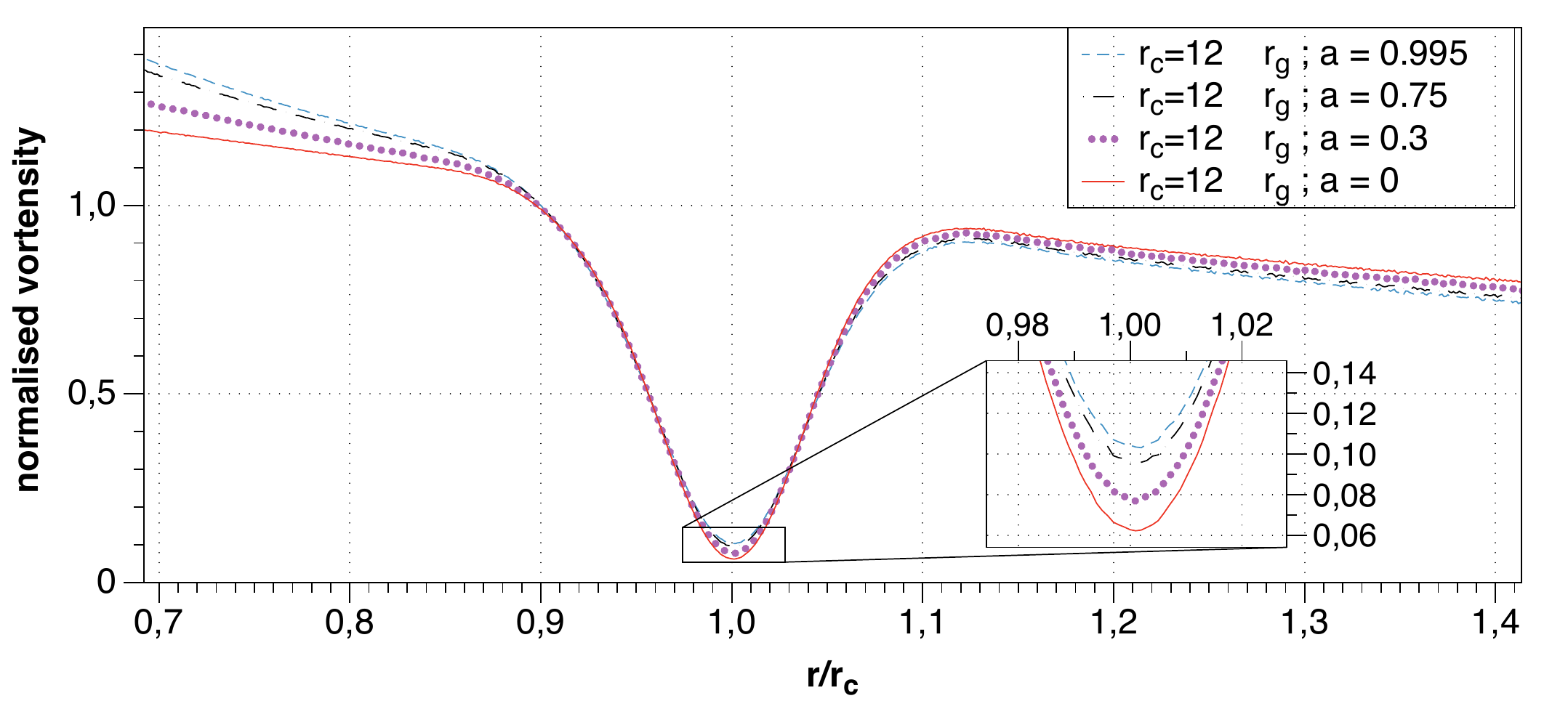}
\caption{{\bf Lower panel}: Radial profile of the vortensity of a disc with $r_c=12 r_g$ for various spin parameters. {\bf Upper panel}: Radial profile of the vortensity of a disc orbiting around a fast rotating disc $a=0.995$ for various corotation radii. All profiles are normalised to their value at the corotation radius  $r_c$ in the case where no bump is present.}
\label{fig:Criter1}
\end{figure}

 \refe{Previous studies \cite[see e.g.][]{Lovelace99} have shown that the RWI is triggered if a function ${\cal L}$ of the gas exhibits a local extremum. Extrapolating its definition from non-relativistic studies, one can define this function (often called vortensity)   as the ratio of the component of the velocity curl perpendicular to the disc mid plane to the surface density, namely
 \begin{equation}
{\cal L}(r) = \frac{\varepsilon^{\theta\rm i}_{\rm j}{\rm v}^j_{;i}}{\sigma(r)}  \label{eq:L}
\end{equation} 
where $\varepsilon^{\rm ij}_{\rm k}$ is a Levi-Civita tensor.} The stability criterion is then linked to the surface density of the disc \refe{$\sigma$} as well as the radial derivative of the rotational velocity of the disc ${\rm v}^{\varphi}$. Designing a density profile such that the disc has a density bump able to fulfill the criteria Eq.\ref{eq:L} leads to the growth of a Rossby wave corotating with the fluid at $r=r_c$. 
  We set the surface density of the disc to
\begin{equation}
\refe{\sigma(r)}= \displaystyle\left(\frac{r_o}{r}\right)^{0.7}\left(1+1.9\exp\left\{-\left(\frac{(r-r_c)}{2\delta^2}\right)^2\right\}\right)
\label{Eq:Densdistri}
\end{equation} 
 where  $r_o$ corresponds to the last stable orbit radius for the fastest spinning black hole we considered ($a=0.995$)
 namely $r_o = 1.34 r_g$. $r_c$ corresponds to the corotation radius of the Rossby waves generated during the computation  and $\delta$ is the typical width of the density bump. The density profile has been chosen so that it {creates a similar criteria for the instability} as one varies both the spin parameter and {the position of the bump}. 
 In order to illustrate such property we display on Fig.\ref{fig:Criter1} the aforementioned instability criterion for various spin parameters and corotation radius. 
 As one can see, all criteria {have a similar shape and} exhibit {only}  small discrepancies near the corotation radius,  leading to {similar enough} disc configurations {allowing us} 
 to compare the growth of the instability for {all} spin parameters.

\begin{figure}
\includegraphics[width=0.5\textwidth]{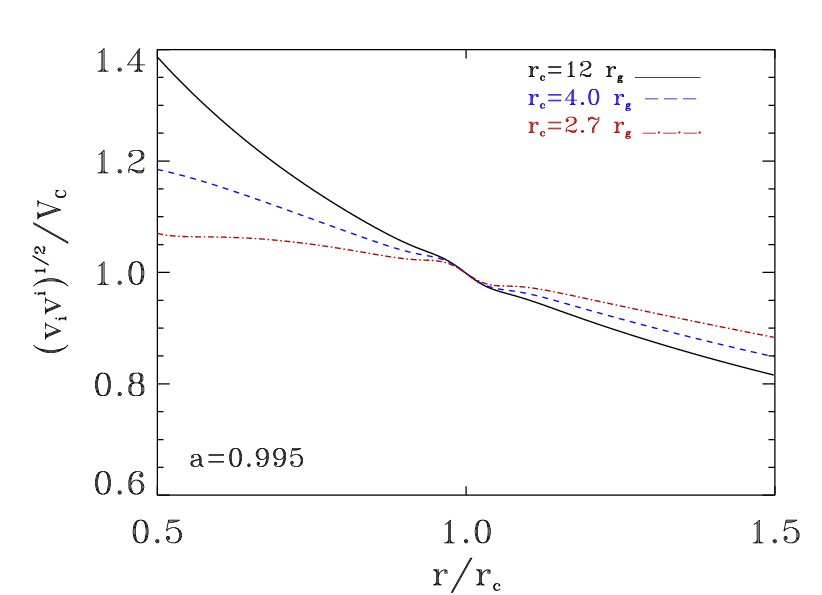}
\includegraphics[width=0.5\textwidth]{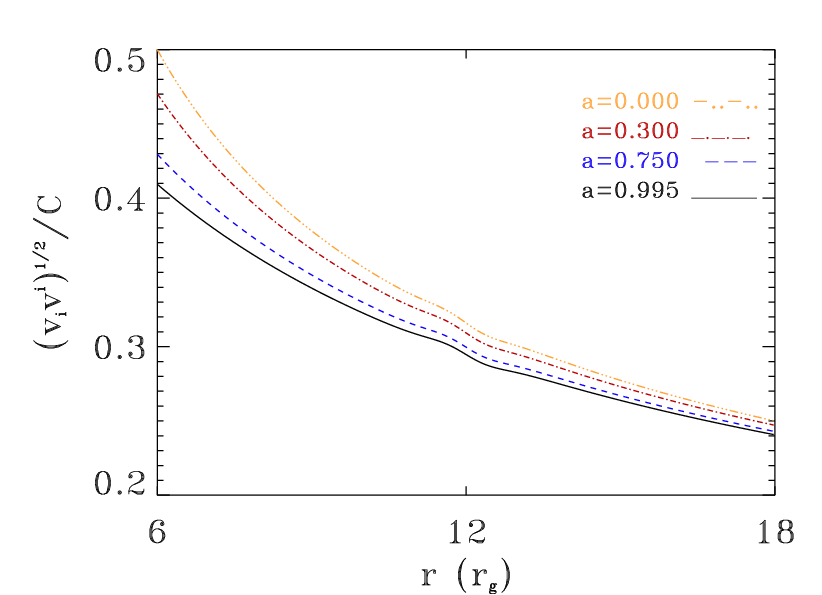}
\caption{\refe{{\bf Lower panel}: Radial profiles of the rotational velocity of discs with $r_c=12 r_g$ for various spin parameters. {\bf Upper panel}: Radial profiles of the rotational velocity of a disc orbiting around a fast rotating disc $a=0.995$ for various corotation radii. All profiles in the upper panel are normalised to their value $V_c$ at the corotation radius  $r_c$.}}
\label{fig_Vel}
\end{figure}

  The radial equilibrium of the disc is achieved by ensuring that the centrifugal force balances exactly the gravitational acceleration induced by the black hole \refe{as well as the pressure gradient}. Such equilibrium is obtained by solving numerically the radial momentum conservation, namely
 \begin{equation}
 {\rm v}^{\varphi}=c\displaystyle\frac{-\partial_r\beta^\varphi\pm\sqrt{(\partial_r\beta^{\varphi})^2+2\alpha\partial_r\gamma_{\varphi\varphi}(\partial_r\alpha+\alpha\frac{\partial_rP}{\rho hW^2})}}{\alpha\gamma^{\varphi\varphi}\partial_r\gamma_{\varphi\varphi}} 
 \label{Eq:shear}
 \end{equation}
 The negative sign corresponds to a counter-rotating disc with respect to the black hole rotation while the positive one stands for a corotating disc. We set the thermal pressure parameters to the same value as in \citet{Cass17}, namely $C_o=1.8\times 10^{-4}$ and $\tilde{\gamma}=3$. These values  have been selected in order to correspond to a thin disc setup whose thickness ratio is \refe{$H/r\sim 4\times 10^{-2}$}. \refe{As an example, we display in Fig.\ref{fig_Vel} the radial profile of the initial rotational velocity of the disc for the same configurations as in Fig.\ref{fig:Criter1}.}
 
 To seed the instability we add random perturbations 
 of the radial velocity (originally null) near the bump 
 \begin{equation}
 \delta\text{v}_{r,\rm ij}=\xi_{\text{ij}}\text{v}_\varphi\exp\left(-\left(\frac{r-r_c}{\delta}\right)^2\right)\nonumber
 \end{equation}
 where $\xi_{\text{ij}}$ is a random variable in computational cell $\rm (i,j)$ verifying $<\xi>=0$ and $<\xi^2>=10^{-10}$ when averaged over the whole computational domain. 

 %%%%%%%%%%%%%%%%
%%%%%%%%%%%%%%%
%%%%%%%%%%%%%%%%
%%%%%%%%%%%%%%%

{In the rest of the paper we will be performing simulations for a wide range of spin but showing only a restricted number of plots. Only Fig.\ref{fig:F10} has all the simulations shown as it 
summarizes our work. Our chosen spins are: $a \in [0, 0.3, 0.5, 0.75, 0.9, 0.95, 0.98, 0.99, 0.995]$}.
 In order to compare the RWI evolution at {so many} different locations in the disc, one has to design computational domains taking into account the different timescales involved in the dynamics of the disc. These timescales are determined by the rotation period of the gas around the black hole denoted as $T(r)$.

 In all our simulations we will keep a constant aspect ratio regarding the radial extension of the computational domain so that it ranges from $r_c/2$ to $3r_c/2$. Such choice ensures that the perturbations are crossing the computational domain over the same time when measured in corotation period $T(r_c)=T_c$ unit. This also enables us to monitor the evolution of the instability and to assess modifications induced by  the relativistic gravity of the black hole, especially when the RWI is triggered close to the black hole.  
 Let us mention that the spatial resolution of the 2D grid is $196\times 600$ for each simulation with the toroidal coordinate $\varphi \in [0,2\pi]$.
 The inner radial boundary \refe{is designed so that the values of physical quantities in the ghost cells are copied from the innermost cells of the computational domain. The outer radial boundary follows the same procedure while azimuthal boundaries are periodic. }
 
 \refe{When triggering the RWI, density and velocity perturbations propagates in both inward and outward directions. Reflection of such perturbations  onto radial boundaries may distort the development of the instability the disc.  In order to monitor such potential distortion we ran several simulations with an increased radial extension of the computational domain. Increasing the radial extent of the domain by a factor $1.5$ has led to  similar saturation levels of the instability in all our tests. As the crossing time of the perturbations is modified in that context we are confident that such reflected waves do not significantly interfere with the growth and saturation of the RWI.}
 In order to follow the temporal evolution of the set of equations presented in the previous section, we used a Harten, Lax and van Leer (HLL) solver linked to a Koren slope limiter. A typical simulation requires several tens of thousands time steps.\\

%%%%%%%%%%%%%%%%%

\section{The RWI beyond the ISCO of Schwarzschild black holes}

%%%%%%%%%%%%%%%%%%%%%%%%%%
\begin{figure*}
\centering
\includegraphics[width=0.32\textwidth]{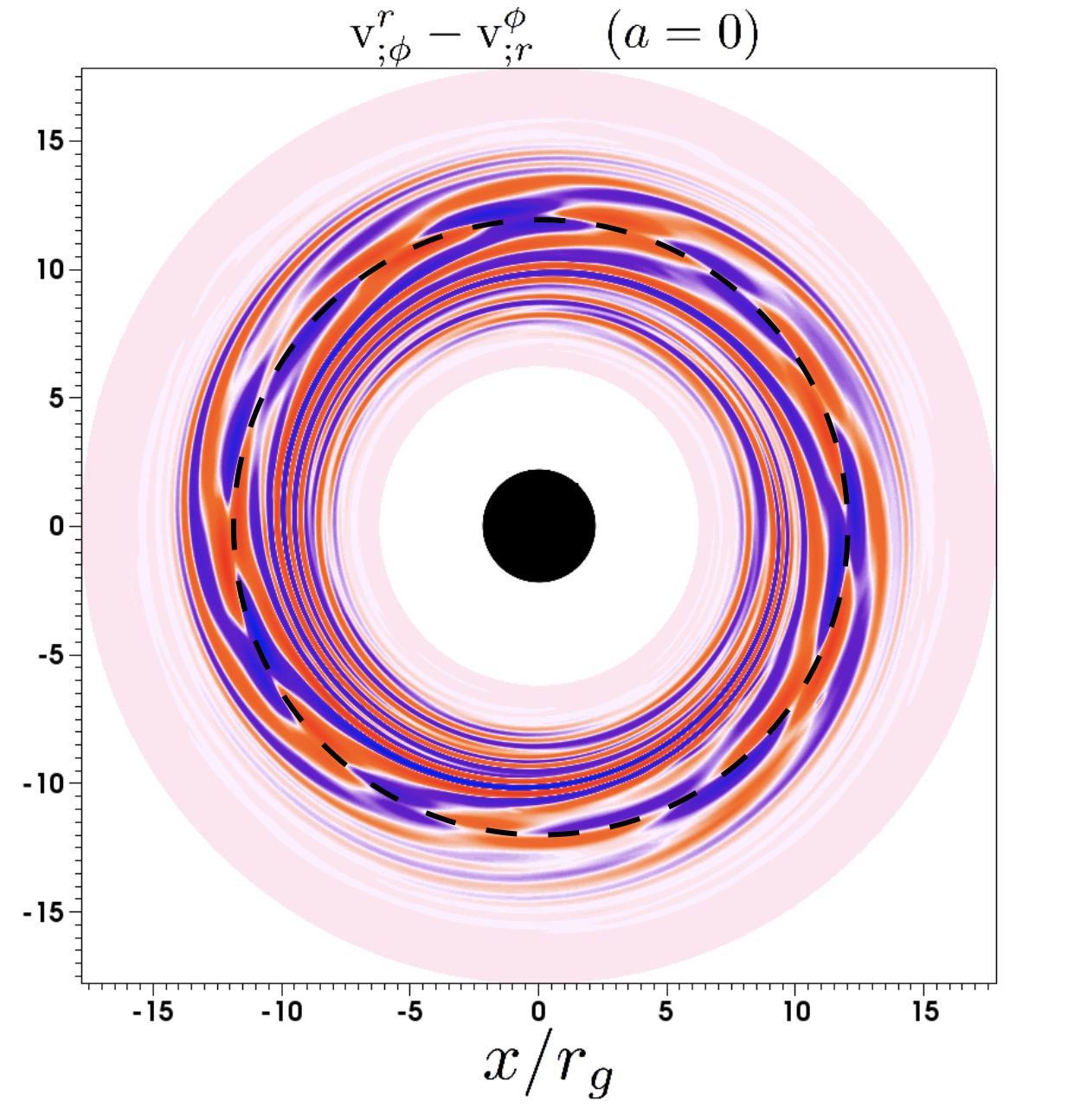}
\includegraphics[width=0.32\textwidth]{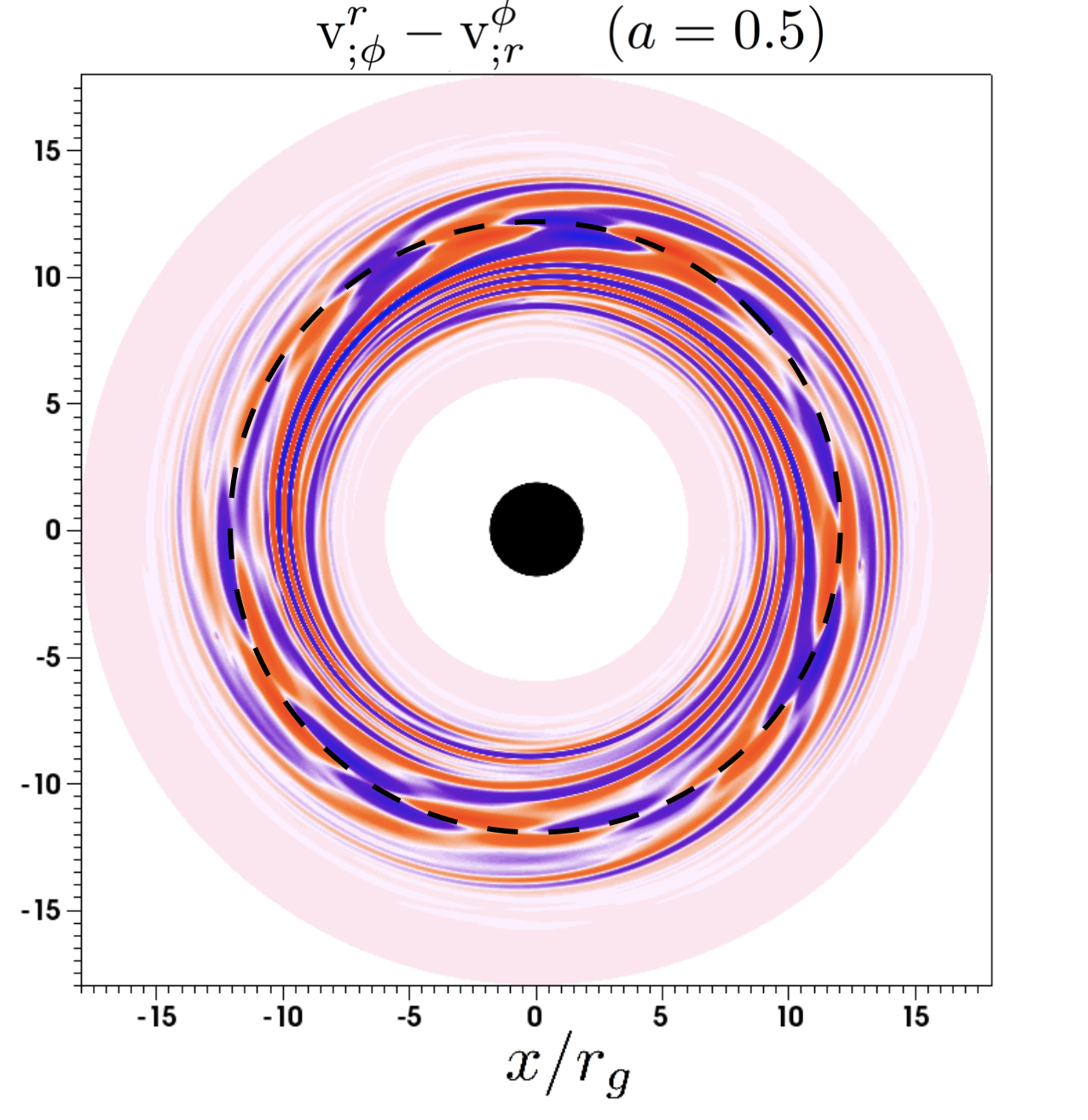}
\includegraphics[width=0.32\textwidth]{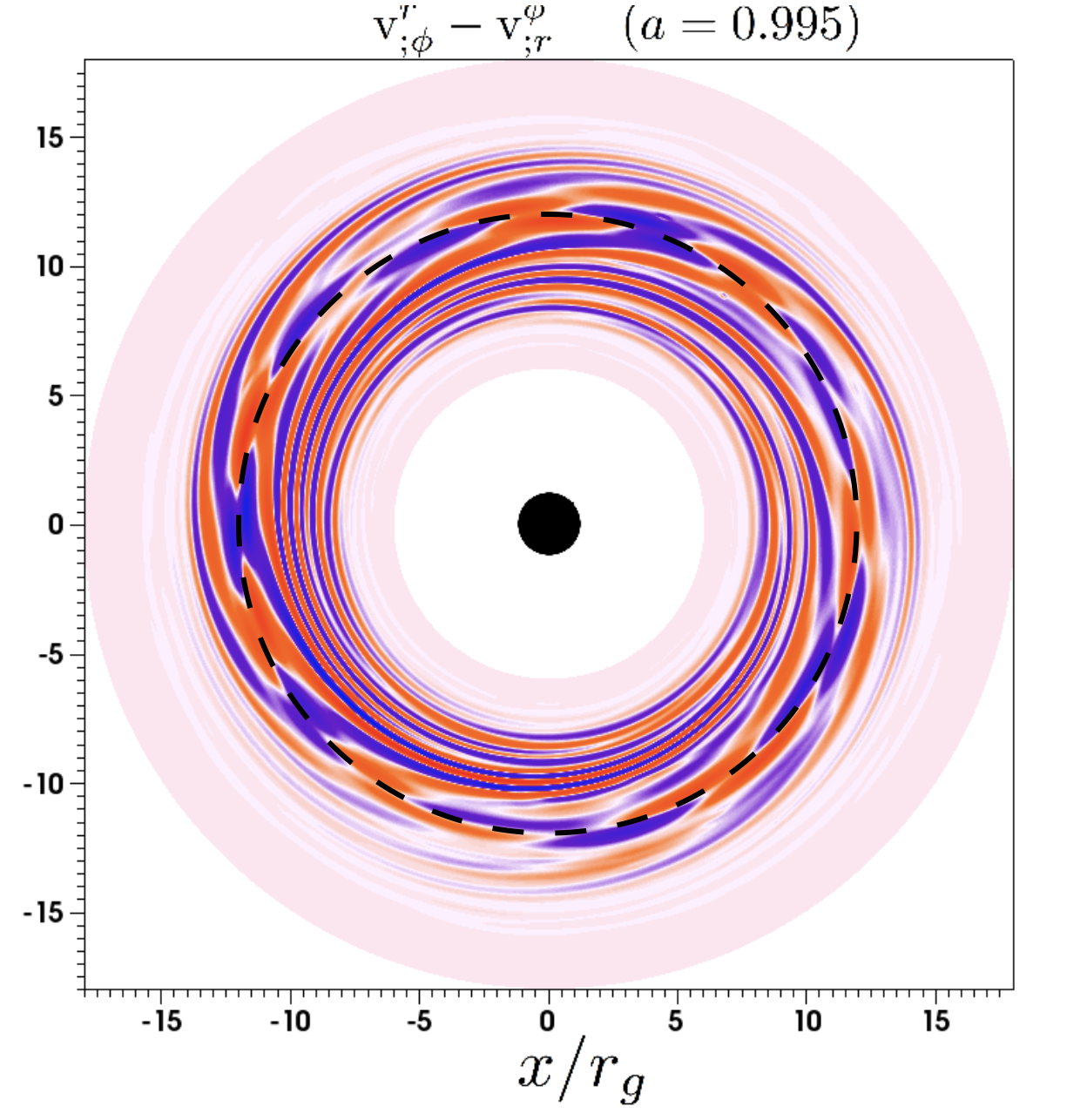}
\caption{Colormap of  $\text{v}^r_{;\varphi}-\text{v}^\varphi_{;r}$ in discs orbiting around three different spinning black holes. All colormaps are performed at the same remote observer time (also called Newtonian time) corresponding to $3T_{\rm c}$ ($T_{\rm c}$ is the disc period at the corotation radius), namely during the exponential growth phase of the instability. \refe{The filled dark circle stands for the size of the event horizon of the black hole while the dashed black line indicates the location of the corotation radius $r_c$.} The three plots 
are very similar which indicates that the RWI can develop in accretion disc at $r_c=12 r_g$ whatever the spin of the black hole. They also support the fact that the spin of the black hole has no noticeable influence on the growth rate of RWI at that distance from the compact object.}
\label{fig:F3}
\end{figure*}
%%%%%%%%%%%%%%%%%%%%%%%%
\begin{figure*}
\centering
\includegraphics[width=0.49\textwidth]{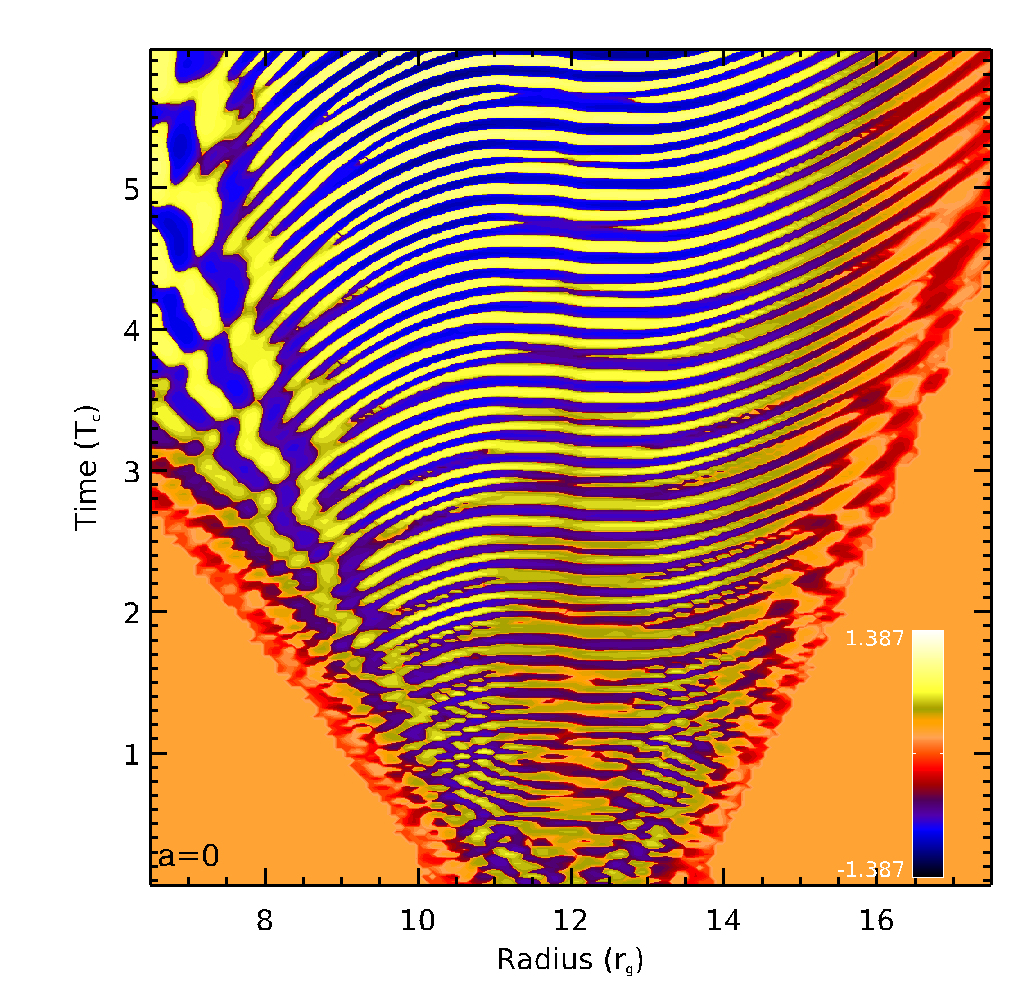}
\includegraphics[width=0.485\textwidth]{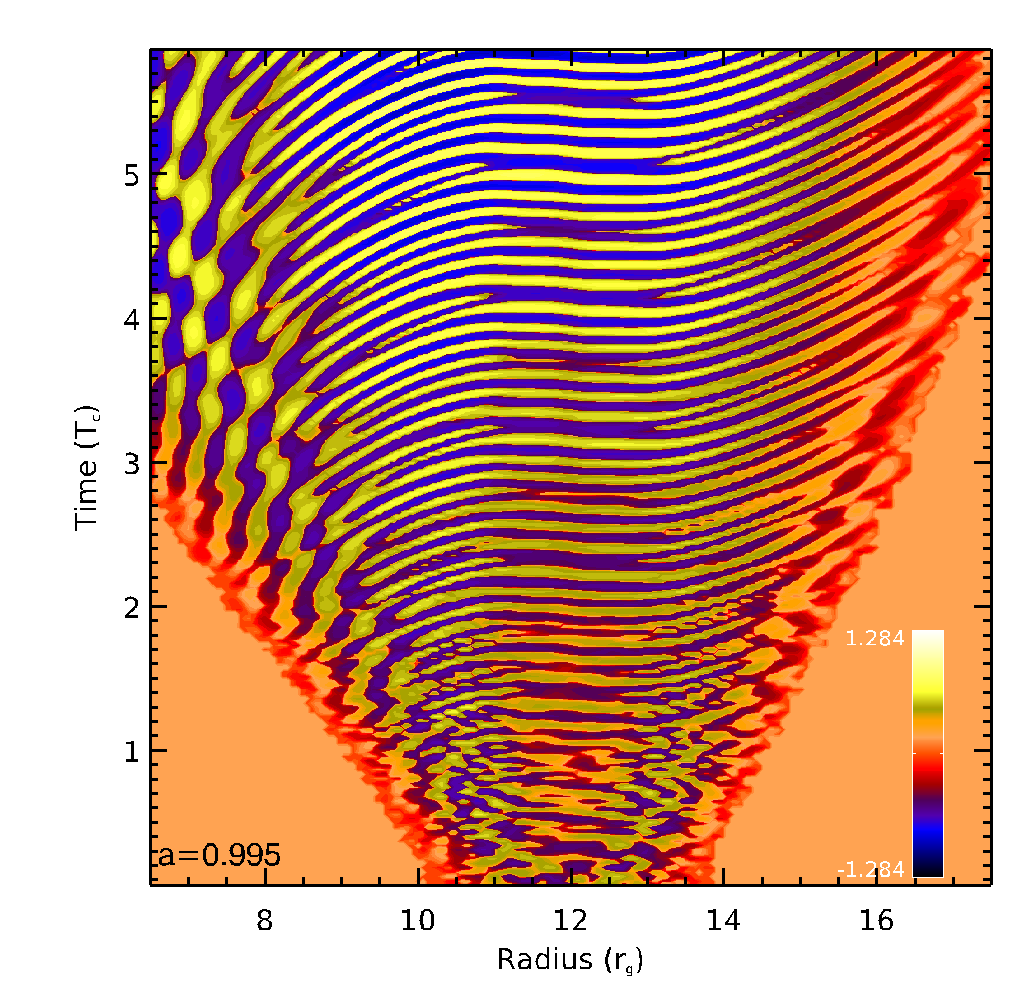}
\caption{Colormap of logarithmic density perturbations as a function of the radius \refe{and Newtonian time measured in disc period units $T_c(r_c)$  at $\varphi_0=\pi$ and with $r_c=12 r_g$} . The left panel corresponds to a disc orbiting around a Schwarzschild black hole while the right one stands for a disc near fast spinning compact object ($a=0.995$). Both plots exhibit the same perturbation propagating front hence reaching the borders of the simulations at the same time. Minor differences appear in the spiral pattern near the inner edge of the simulation as the shearing of the disc exhibits differences as can be seen on Fig.\ref{fig:Criter1}.}
\label{fig:F5}
\end{figure*}
%%%%%%%%%%%
%%%%%%%%%%%%%%%%%%%%%%%%%%%%%
\begin{figure}
\includegraphics[width=0.5\textwidth]{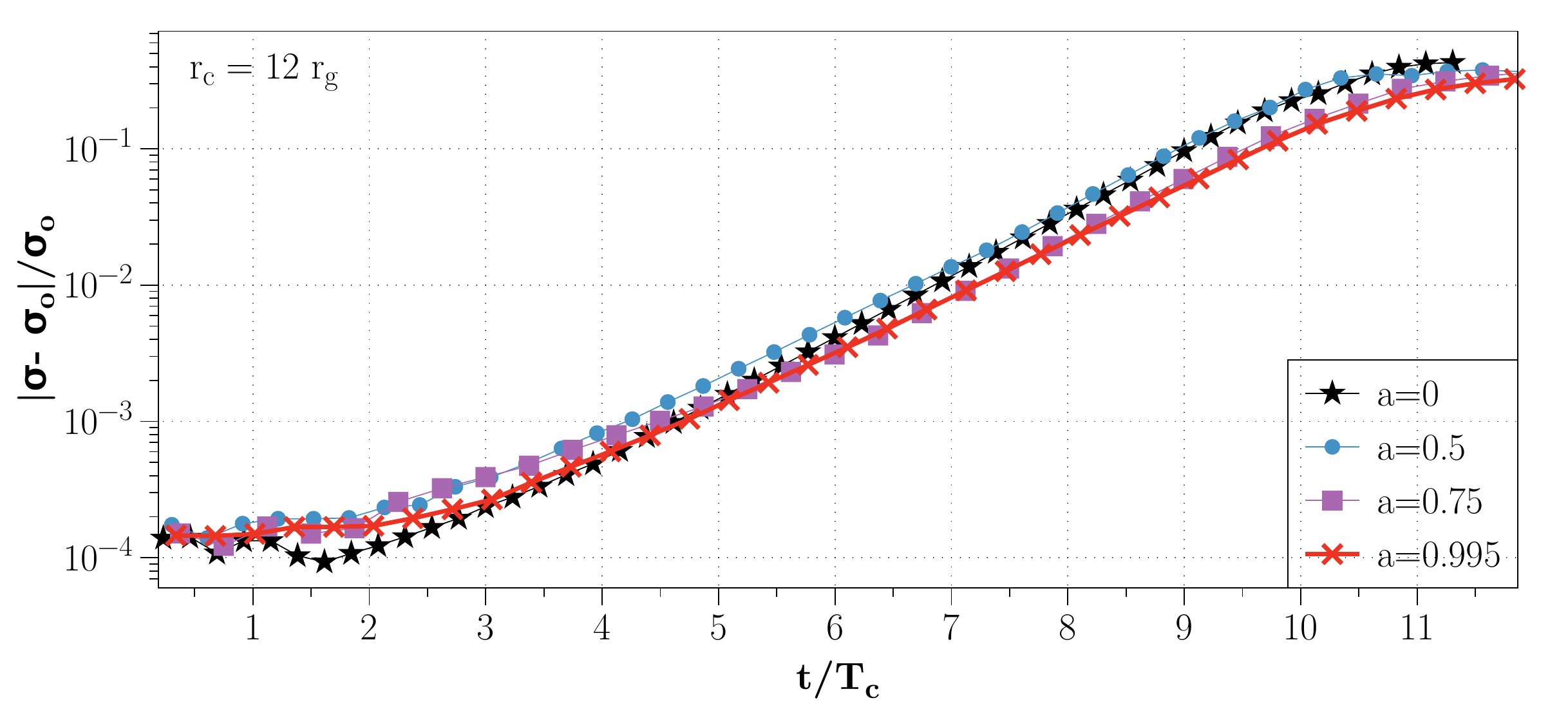}
\caption{Temporal behaviour of \refe{normalised maximal density fluctuations occurring in the vicinity of the density bump with respect to Newtonian time ($\sigma_o$ being the density of the disc at $r=r_c$)}. The various curves exhibit similar shapes even though low spin black hole cases exhibit slightly larger growth rates of the instability than in higher spin simulations. }
\label{fig:F2}
\end{figure}
%%%%%%%%%%%%%%%%%%%%%%%%%%

  The aim of this section is two fold, first reproduce the results from our previous Schwarzschild paper at $r_c=12 r_g$ {to ensure that we are following the RWI} and {then } show what is the impact of different spins at that one location.  Indeed, even if the impact of the spin is small at $12 r_g$ the velocity profiles are different and it is interesting to check if we could find observable differences between spins even when the inner edge of the disc is not at its last stable orbit.
 
\subsection{Confirming the presence of the RWI}

{The first thing that we are looking at is to confirm that we are actually probing the RWI in our simulations. This is the reason we started with the $12r_g$ cases that was 
already studied in a previous publication and proven to be the same as the Newtonian RWI.} In \citet{Cass17} we investigated the influence of the location  of the RWI with respect to the last stable orbit around a Schwarzschild black hole. Considering fast rotating black holes, the velocity profile of the  accretion disc near the ISCO of Schwarzschild black hole differs from the non-spinning case as stated by Eq.(\ref{Eq:shear}) \refe{and shown in Fig.\ref{fig_Vel}}.  In order to  assess the influence of such velocity shift we have performed various simulations of the RWI at the very same position, namely $r_c=12r_g$ around different rotating black holes. We have considered accretion discs orbiting around black holes whose spin ranges from the Schwarzschild case up to $a=0.995$, namely fast rotating black holes.
 
A way to investigate the existence of the RWI is to look at the compression and shear waves arising from the bump and propagating both inward and outward in the disc. We have displayed in Fig.\ref{fig:F3} the colormaps of  the perpendicular component of the curl of the disc velocity for three different computations at an early time (colormaps are performed using linear scales). The figure is consistent with the development of a RWI near the density bump in the disc.  Fig.\ref{fig:F3} shows the presence of vortices spreading on both sides of the density maximum. It is noteworthy that the three cases presented in this figure cover the whole span of the spin of Kerr black holes (from $a=0$ to $a=0.995$). As the three simulations exhibit very similar patterns, one can {see} that the growth of the instability is only loosely coupled to the spin of the central object when triggered beyond the ISCO of a Schwarzschild black hole. 

The temporal evolution of the instability can be synthesised when displaying the radial profile of the \refe{density perturbations at a given angle ($\varphi=\pi$) with respect to Newtonian time (measured in $T_c$ unit)}.  We have performed such representation in Fig.\ref{fig:F5} for both the non-spinning and fast spinning black hole cases (corresponding to left and right panels respectively of Fig.\ref{fig:F3}). We notice that density perturbations (along with the velocity ones) are initially injected within the density bump whose maximum is located at $r_c=12\ r_g$. As time goes by, perturbations propagate both inwards and outwards accordingly to the local sound velocity that is fully determined by the radial density profile and the lapse function $\alpha(r,a)$.
As already mentioned in the previous section, we have chosen a density profile 
 so that the ratio of the  sound speed to the rotational velocity is similar in all simulations. 
The global aspect of the perturbations envelope is similar for both cases which is in agreement with such an identical velocity ratio. It is noteworthy that having similar propagating fronts for the perturbations  in the two cases illustrates the fact that  both lapse functions $\alpha(r>6r_g)$ are similar and actually close to unity in the computational domain. At such distance from the black hole, the impact of the spin parameter upon the physics of RWI and the propagation of perturbation waves is minimal. For the sake of completeness, let us mention however that small differences do exist between the two cases. Indeed the local perturbation patterns near the edge of the envelope are somehow different. Even though lapse functions are close to one another they are not identical which leads to different radial gravitational forces and hence different disc rotational velocity as shown by Eq.(\ref{Eq:shear}) \refe{and Fig.\ref{fig_Vel}}. The shearing of velocity perturbation arises from discrepancies between rotational velocity in the two cases and affects the shape of the waves as they travel through the disc.

\subsection{Behaviour of the RWI growth rate beyond the ISCO of Schwarzschild black holes}
Now that we have shown that we are developing the RWI we can compare the actual growth of the instability for spins comprised in the sample $[0, 0.3, 0.5, 0.75, 0.9, 0.95, 0.98, 0.99, 0.995]$.
Having designed an exact disc equilibrium enables us to track the fluid quantity perturbations very accurately and to monitor the growth of the instability. In all computations, we have obtained an exponential growth of the instability whose temporal behaviour is presented in Fig.\ref{fig:F2}. All curves exhibit {a similar} shape despite some minor differences regarding the growth rate of the instability. Indeed, we do observe that low spin cases exhibit a slightly larger RWI growth rate than higher spin  configurations.
This can be understood as the local rotational velocity of the gas is decreasing as the spin of the black hole increases due to the Kerr metric structure. 
The final saturation level of the instability is quite similar in all cases so that we can safely assume that the spin of the central black hole has no influence upon the saturation of the RWI when triggered beyond the ISCO of Schwarzschild black holes.

 \section{RWI {down} the gravity well of spinning black holes}

In this section we address the development of the RWI within an accretion disc orbiting at smaller distance from a spinning black hole than in \citet{Cass17}. As already mentioned, the innermost stable circular orbit radius of the disc  is a decreasing function of the spin parameter of the black hole. Considering high spin black holes is then an opportunity to investigate the RWI deep down the gravity well of Kerr black holes. 

\subsection{ Triggering the RWI closer to spinning black holes}

In order to illustrate the influence of the spin of the black hole upon the development of the RWI, we intend to trigger it at
 smaller and smaller position in term of gravitational radius, the lower bound being dependent on the actual spin studied for each case. Indeed, as we increase the spin we can get closer to the black hole in term of  gravitational radius.  
 %%%%%%%%%%%%%%%
\begin{figure}
\centering
\includegraphics[width=0.497\textwidth]{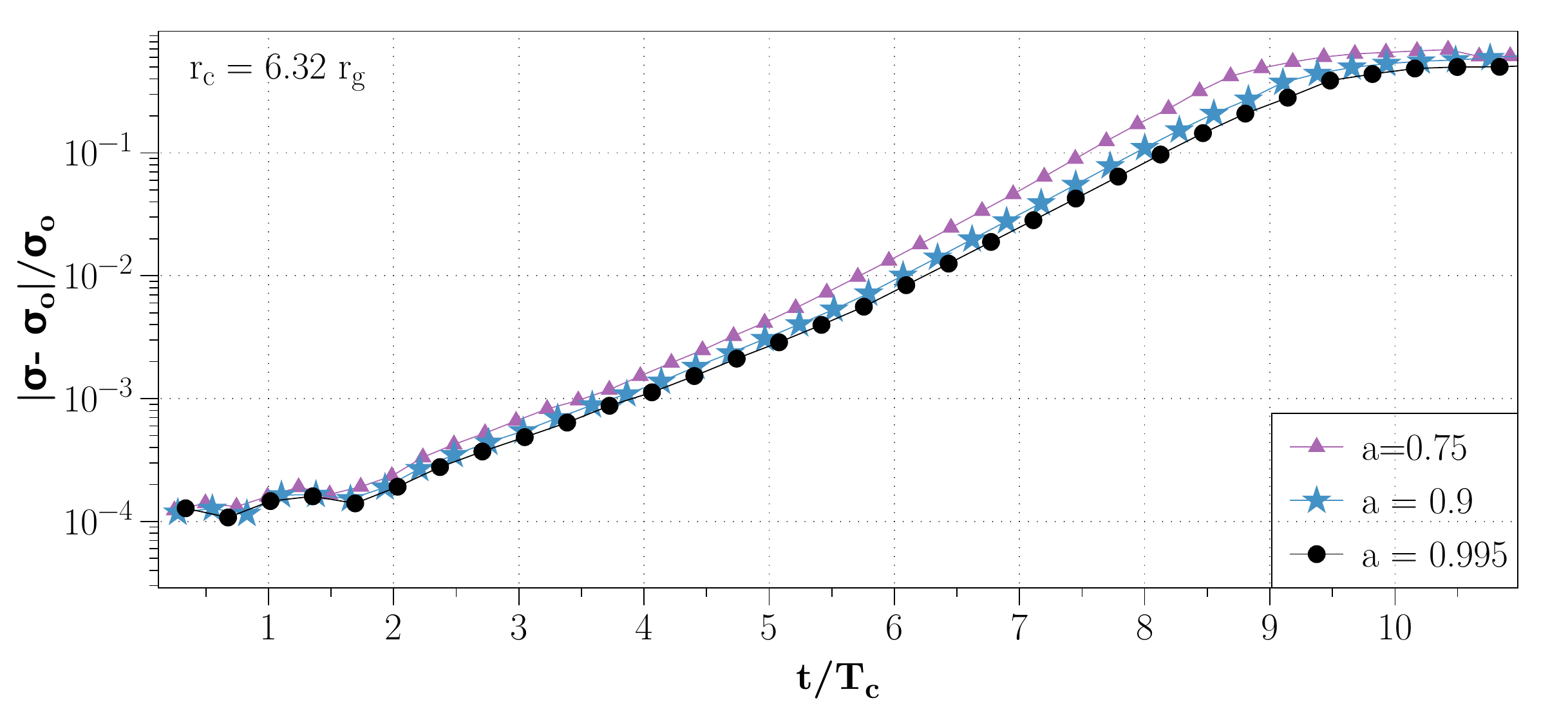}
\caption{ Temporal evolution of the density perturbations induced by a RWI triggered at $r_c=6.32 r_g$ around various fast-spinning black holes with respect to Newtonian time.}
\label{fig:F9}
\end{figure}
%%%%%%%%%%%%%%

  %%%%%%%%%%%%%%%%
 %%%%%%%%%%%%%%%%%
 %%%%%%%%%%%%%
\begin{figure*}
\centering
\includegraphics[width=0.324\textwidth]{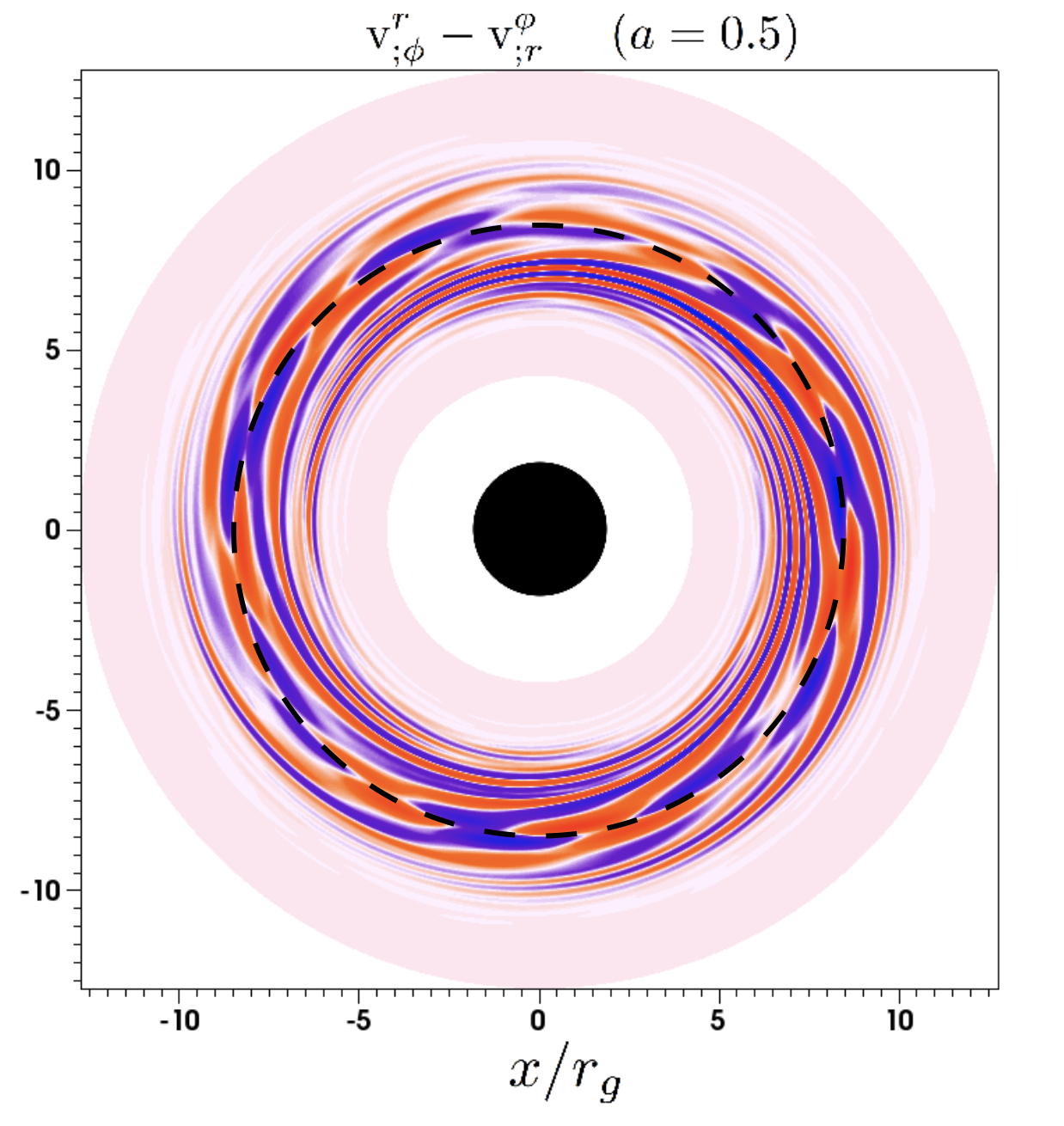}
\includegraphics[width=0.315\textwidth]{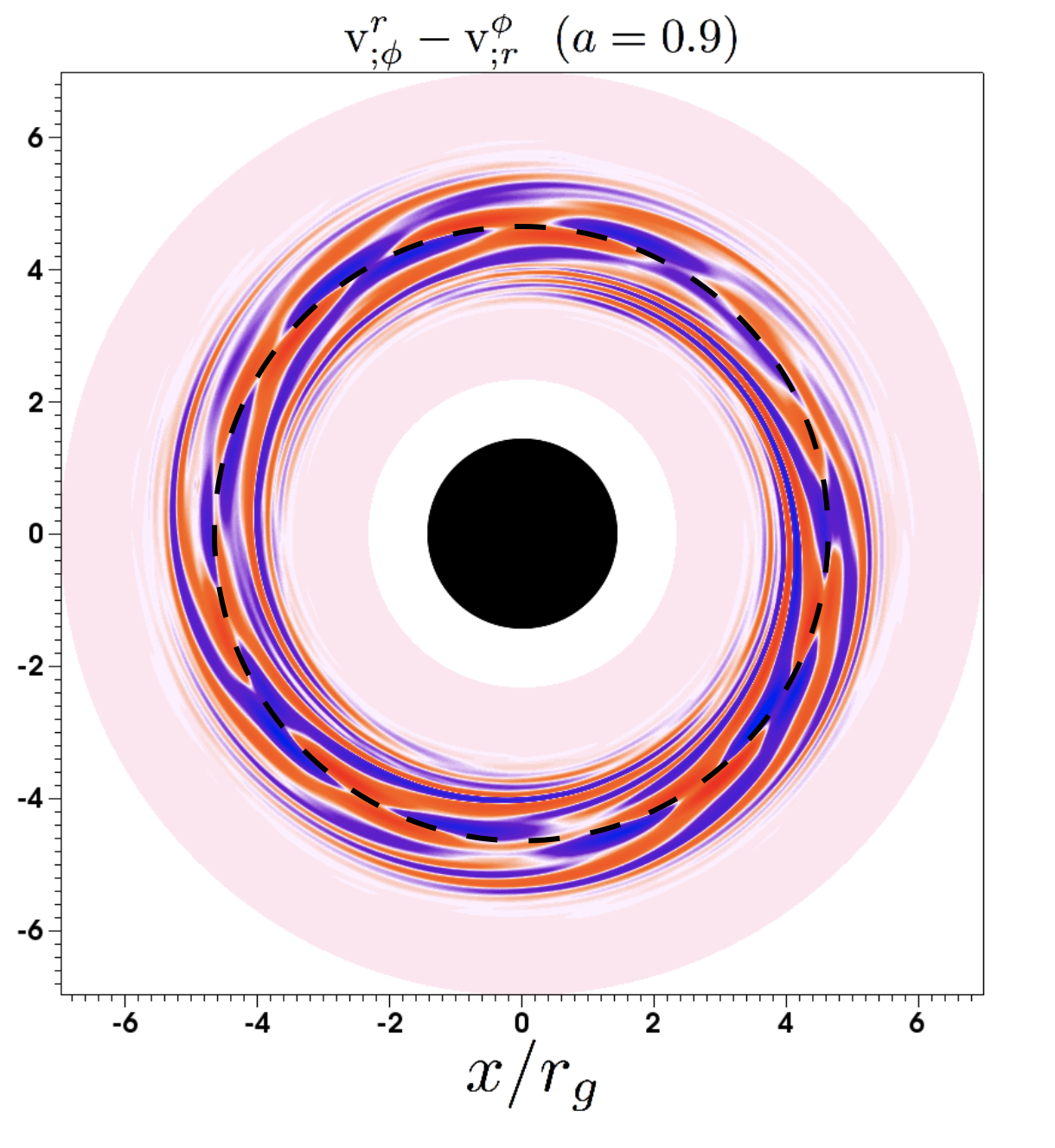}
\includegraphics[width=0.324\textwidth]{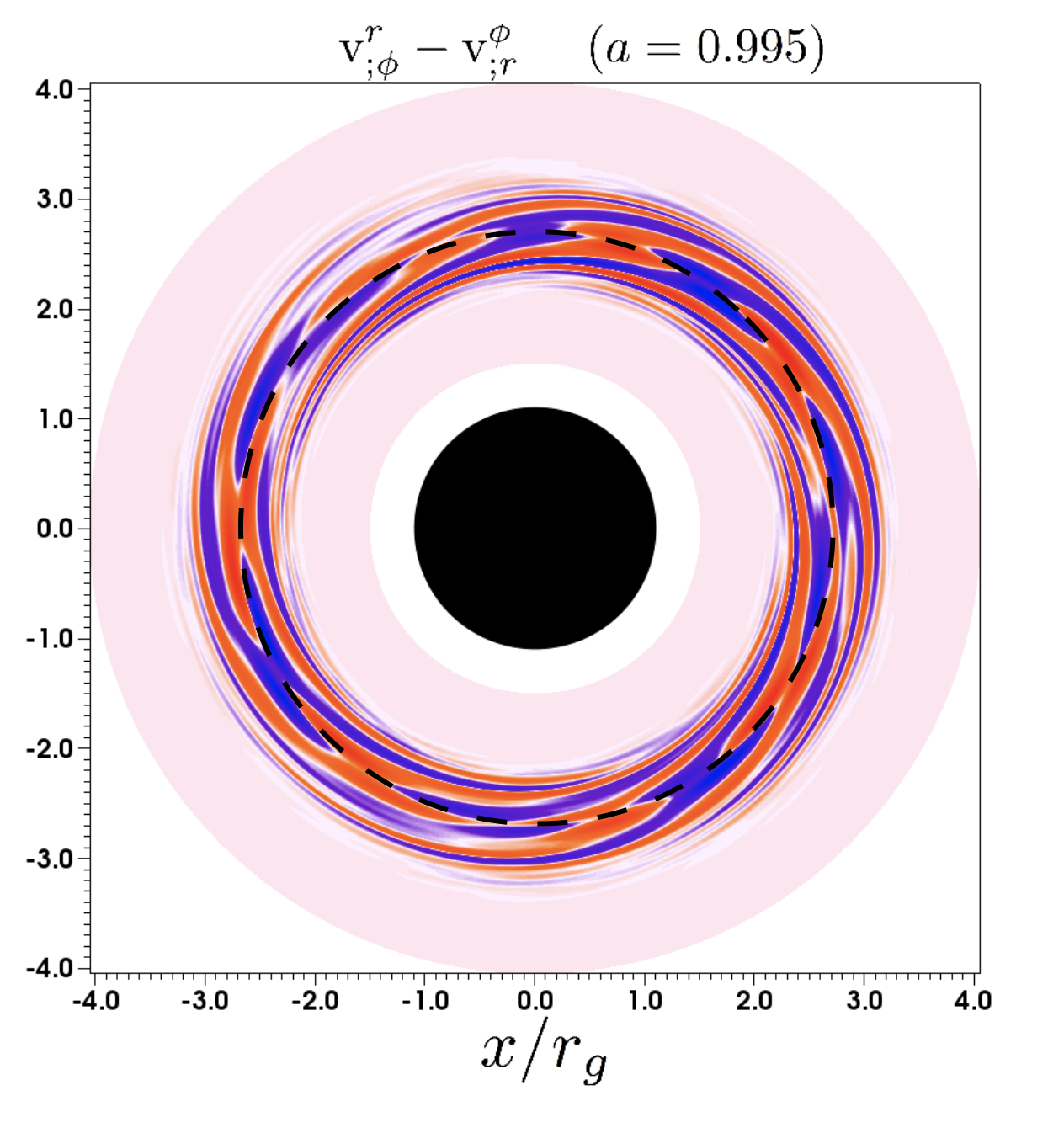}
\caption{Same plots as in Fig.\ref{fig:F3} but for smaller corotation radii, namely  $r_c=8.5, 4.6$ and $2.7 r_g$. We display here the velocity curl for three spin parameters, namely $a=0.5, 0.9$ and $0.995$. The three plots are performed at the same remote observer time corresponding to  $5$ disc rotations of the bump $T_{\rm c}(r_c)$ .  The pattern associated with the development of the RWI within the bump is similar to the ones obtained in Fig.\ref{fig:F3}. }
\label{fig:F6}
\end{figure*}
%%%%%%%%%%%%%%%

\begin{figure}
\centering
\includegraphics[width=0.49\textwidth]{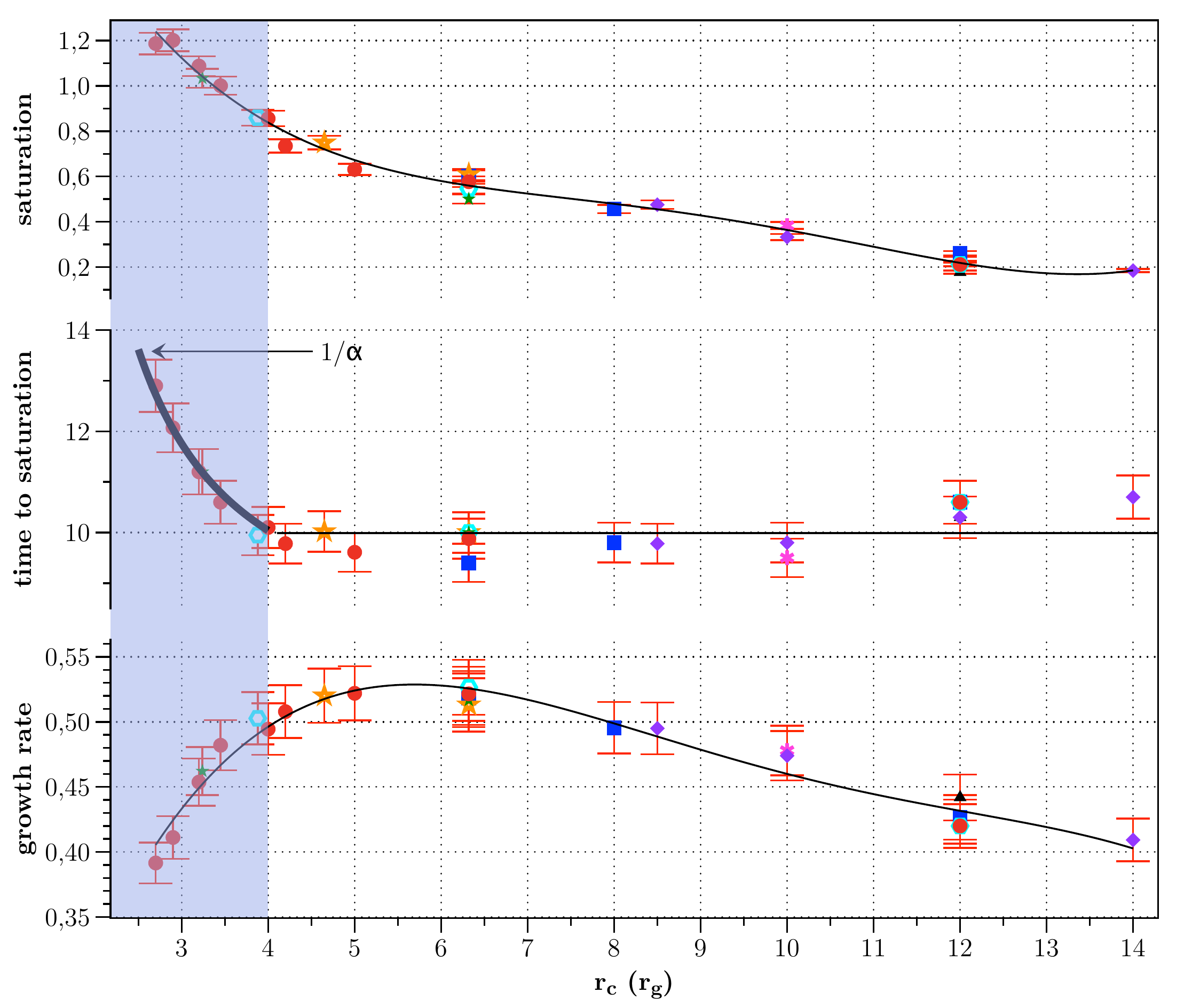}
\caption{ {\bf (Up)} Relative saturation level of the RWI (measured with density variation) in the same cases as the other panels. {\bf (Middle)} Time to saturation of the RWI in corotation period time units $T_c$ for the various simulations (multiple spin parameter $a$ and $r_c/r_{ISCO}$ ratio).{\bf (Bottom)} Linear growth rate of the RWI measured in $T_c^{-1}$ units. 
\refe{Symbols represent the spins as follows: $a=0.9$ an orange hollow star, $a=0.98$ a green star, $a=0.3$ a pink star, $a=0.95$ a cyan hexagon, $a=0.995$ a red circle, $a=0$ a black triangle, $a=0.75$ a blue 
square and $a=0.5$ a purple diamond. The lines are polynomial fits to the data. }}
\label{fig:F10}
\end{figure}

 In the previous section we have presented the growth rates of the RWI at $r_c=12\ r_g$ for various spin parameter ranging from $0$ to $0.995$. 
 In Fig.\ref{fig:F2}  one can see that for spin parameters larger than $0.75$, no significant difference whatsoever can be noticed between the various runs. 
 In order to check if this statement holds for RWI locations smaller than $12\ r_g$ we first have performed similar simulations at a smaller corotation radius while preserving a sufficiently large span of spin parameters. In that respect we used $r_c=6.32\ r_g$ which correspond to the inner edge of the computation box at the ISCO of the $a=0.75$ case, hence
 giving us a sufficiently large dynamical range of spin. 
 The corresponding growth rates are displayed in Fig.\ref{fig:F9} and one can easily see that all growth rates coincide. Let us mention that we performed numerous simulations considering such $r_c$ but only displayed the outcome from three of them for the sake of clarity (see Fig.\ref{fig:F10}. for a summary of all simulations).
  We can then {conclude} that at {this} location from the center of the black hole, no significant spin induced effect is visible upon the development of the RWI. This was actually expected as the lapse function $\alpha$ of the Kerr metric remains almost flat over the computational domain of these simulations when varying the spin parameter. This in turns leads to  similar rotational velocity distributions of the discs as well as a minimal temporal distortion between the inner and outer boundary of the simulation.
\newline

 In order to push forward our investigation of the potential influence of the spin of the black hole upon the RWI, we have performed many simulations at closer locations to the black hole. We have selected three of them probing smaller and smaller corotation radii  deep down the gravity well of increasingly rotating black holes. The selected simulations out of the numerous computations we have performed consider RWI triggered at corotation radii $r_c=8.5r_g, 4.6 r_g$ and $2.7 r_g$ around black holes whose spin parameter is $0.5$, $0.9$ and $0.995$ respectively.  
   {Similarly to the simulations displayed in Fig.\ref{fig:F3} we show on}  Fig.\ref{fig:F6} {the} colormaps representing the curl of velocity perturbations relative to the selected simulations.
 The left panel of this figure shows the $a=0.5$ case during the exponential growth phase of the simulation where vortices induced by RWI have already formed and where spiral waves have started to propagate both inward and outward.  In such case we obviously recover the usual  behaviour of the RWI as presented in Sect. 3 and 4.1 and in \citet{Cass17} .

The middle and left panels of Fig.\ref{fig:F6}  exhibit the curl of velocity perturbations for  RWI triggered closer to the compact object, namely at $r_c=4.6 r_g$ and $r_c=2.7r_g$. 
Looking at middle and right panels of Fig.\ref{fig:F6}
we see that, for those early time, the RWI develops in a similar fashion as in previous simulations. Indeed, we do observe the vortices induced by the RWI development altogether with the creation of spiral waves arising from the propagation of perturbations in both directions.

\subsection{Impact of the time dilatation and major consequence of high spin}

\begin{figure*}
\centering
\includegraphics[width=0.49\textwidth]{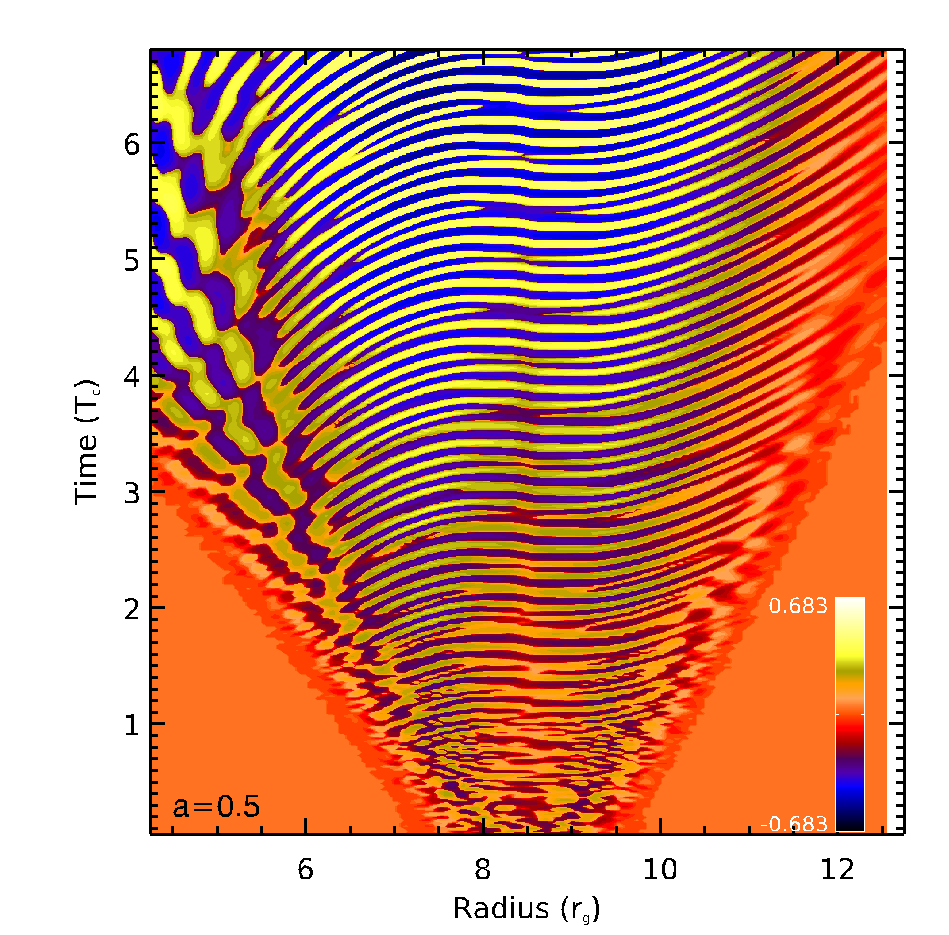}
\includegraphics[width=0.485\textwidth]{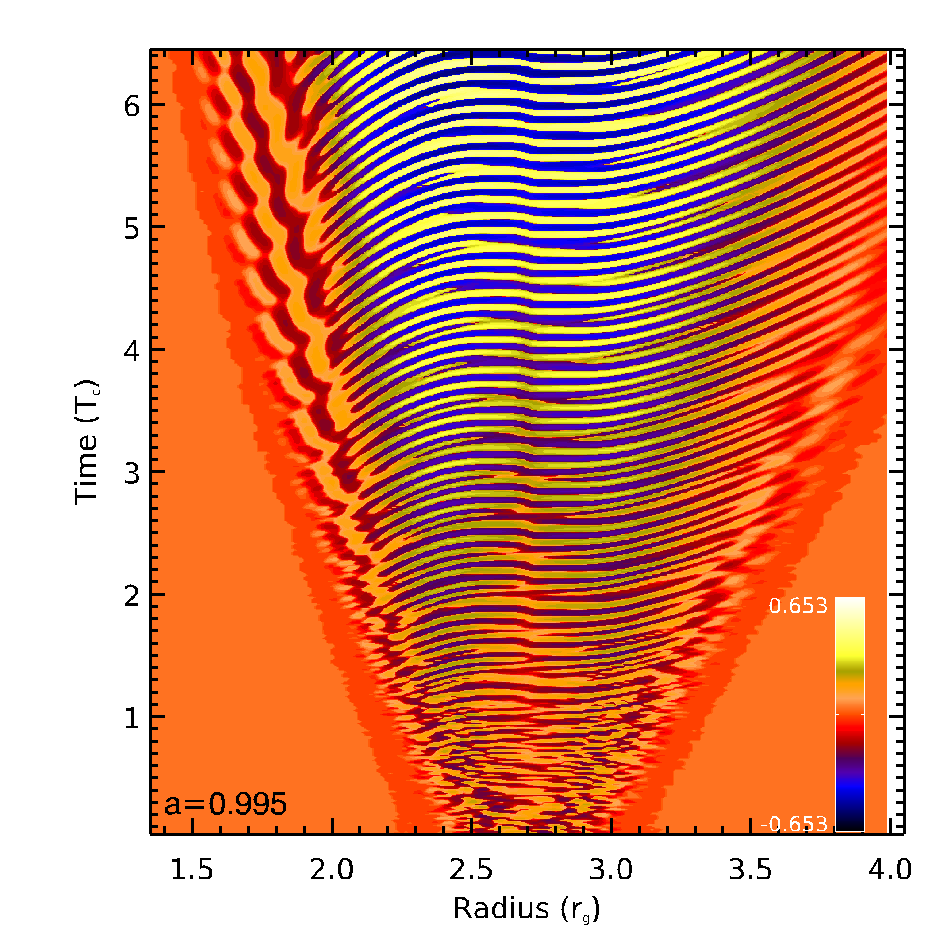}
\caption{Same plots as in Fig.\ref{fig:F5} but for smaller corotation radii, namely  $r_c=8.5r_g$ and $2.7 r_g$. These temporal evolution plots correspond to left and right simulations presented in Fig.\ref{fig:F6}. While the left panel recovers the same patterns as Fig.\ref{fig:F5}, the right panel, corresponding to the smallest corotation radius considered, displays a much slower inwards propagation of the spiral waves. This  effect stems from clock rate dilatation induced by the relativistic gravitational field of the black hole since clock rate is getting smaller as one approaches the event horizon (located at $r_{\rm EH}\sim 1.1 r_g$). }
\label{fig:F8}
\end{figure*}

 The curl of velocity perturbations is displayed using a linear colortable so it is not easy to see if the propagation of these perturbations is affected by the local gravitational field. 
  In order to study the impact of time dilatation in later stage of the RWI evolution it is easier to look at the colormap of the density perturbation as a function of radius and Newtonian time \refe{(in unit of $T_c$)}. 
 On the left panel of Fig.\ref{fig:F8}, corresponding to the $a=0.5$ case, the lower spin pattern is  recovered with both the inwards and outwards propagation of the waves toward the edges of the simulation domain. 
 On the right panel of Fig.\ref{fig:F8} we can see that the inner edge of the wave propagating envelope is slightly bent  as the inwardly propagating perturbations have not reached the inner border of the simulation after more than $6 T_c$ whereas in the $a=0.5$ case the waves have reached the border after $3.5T_c$. 
 In the frame of a remote observer  the propagation of the waves will then appear to be delayed. This effect is due to the clock rate dilatation induced by the strong gravity of the fast spinning black hole leading to  an apparent reduced sonic velocity in the disc, namely $\alpha c_S$. Indeed in the local Eulerian frame, the radial velocity of the perturbations is $c_S$ so that the physics of the instability remains inherently the same in the local frame but its perception from a remote location distorts the global picture of the instability as the various parts of the disc do not experience the same time clock rate.
 In the end, such distortion leads to a new balance between inwards and outwards instability modes.
 
%%%%%%%%%%%%%
%%%%%%%%%%%%%%% 
 %%%%%%%%%%%%%%%%%%%%%
%%%%%%%%%%%%%%%%%%%%%%%%%%%%%%%%%%%%%%

%%%%%%%%%%%%%%%%%%%%%%%%%%%%%%%%%%%%%%%%
\subsection{Growth rate and saturation level of the RWI around Kerr black holes}

{In order to study the impact of the spin on the RWI we}
 have performed numerous simulations considering corotation radii ranging from $2.7 r_g$ up to $14 r_g$ and black hole spin parameter spanning from zero up to $0.995$. We have measured for all simulations the linear growth rate of the instability as well as the saturation level and the time needed to reach the non-linear saturated stage of the RWI. {Those are
 summarized in  Fig.\ref{fig:F10} which expands on the results presented before using as parameters of the instability its  growth rate, its saturation level and the time taken to reach this saturation.}
 
  {Indeed, the first thing we see is that for the}  RWI occurring at a given position around a rotating black hole {it} leads to 
  relatively similar values for instability parameters whatever the spin of the black hole.  
{ The immediate consequences of this is that} at a given position{, hence a given frequency of variability,} all spins are indistinguishable. This finding undoubtedly proves that the growth of the RWI is inherently a local physical mechanism independent of the the value of the spin. This is also sustained by the fact that the corotation radii where the most dispersion is observed correspond to locations where the criteria of the RWI, with our setup, has the most disparity. This proves that local physical conditions control the growth of the instability (Fig.\ref{fig:Criter1}).  
While Fig.\ref{fig:F10} implies that spins are indistinguishable regarding RWI at a given location, it also shows that the properties of the RWI differs from one location to another. As the smallest corotation radius can only be met around high spin black holes, we then observe an indirect influence of the spin of the black hole upon the characteristics of the RWI through the time clock dilatation that the waves experiences when propagating toward the inner edge of the disc.
  \newline
   
 If we look more closely as the lower panel of Fig.\ref{fig:F10}, which displays the growth rate of the instability measured in $T_c^{-1}$ units, 
 we see that it goes through a maximum near $r_c=5 r_g$.  At first this result might seems unintuitive  as one would expect the growth of the instability to be stronger as the gravity induced by the black hole gets more intense. In order to explain such behaviour one have to keep in mind that the criterion for the RWI, hence ultimately its growth rate, depends on the rotational velocity of the disc ${\rm v}^{\varphi}$ which is smaller for high spin than in low spin cases essentially because of the rotational shift of the spacetime induced by the rotation of the black hole. Mathematically this is translated by the presence of both a non-vanishing shift vector $\beta^{\varphi}$ and a flatter lapse function $\alpha$ leading to a reduced local rotational velocity of the fluid. As the shift vector is a rapidly varying function tending toward zero, it is then expected to play a role only at short distance from the black hole as we observe here.

The middle panel of Fig.\ref{fig:F10} represents the time needed for the RWI to reach its saturation level \refe{in units of $T_c$}. For corotation radii greater than $4 r_g$ we do observe that the saturation time is roughly constant whatever the spin parameter we considered for the simulation. Most of the differences are related to the difference in the criterion coming from the initial condition that we could not perfectly match for highly different spins. 
However all simulations having a corotation radius smaller than $4r_g$ exhibit larger saturation time not compatible with only the dispersion in initial criteria. 
We have noticed that the increase is actually scaling as the inverse of the lapse function measured at the corotation radius. This finding  is the result of the time clock rate distortion mentioned in the previous subsection where the inward wave appears to propagate at a smaller velocity for a corotation associated observer \refe{(i.e. Zero Angular Momentum Observer)}. The outcome of this distortion is that the saturation time increases as $1/\alpha(r_c)$ compared to simulations where the value of the  lapse function remains flat and close to unity over the whole computational domain, namely simulations having $r_c > 4r_g$.

Lastly, the upper panel of Fig.\ref{fig:F10} stands for the saturation level of the instability in terms of relative density perturbation compared to the initial value of the disc density. We clearly see that the saturation level is increasing as the corotation radius is decreasing. This behaviour is easily understood when the corotation radius is greater than   $4 r_g$ as the time
to reach saturation is roughly constant for an increasing growth rate. It gets interesting when we arrive in the domain where the lapse function impacts the development of the 
instability. Indeed, while we have a decreasing growth rate, the time to reach saturation increases even more, leading to an overall increase of the saturation level.

%%%%%%%%%%%%%%%%%%%%%%%%%%
%%%%%%%%%%%%%%%%%%%%%%%%%%
%%%%%%%%%%%%%%%%%%%%%%%%%%
\section{Conclusions and perspectives}

The RWI has been proposed as a model for different kind of variability occurring in black-hole system but those works were limited to pseudo-Newtonian and 
Schwarzschild metric.
Using our new GR-code  \citep[presented in][]{Cass17} we performed the first simulation of the RWI in a Kerr metric, demonstrating not only its existence (up to a spin of 
$0.995$) but also exploring how its behaviour is modified with spin.

As a matter of fact, while we have shown that the local behaviour of the instability  follows the same general trend from the Newtonian case to a high spin in the Kerr metric, 
we have also found
some distinctions in the behaviour when the instability develops closer than four $r_g$ of the black hole. In order for the disc to reach this radius the spin needs to be high (it corresponds
to the last stable orbit for a spin of $a=0.57$) and the effect of the lapse function become non-negligible.

Indeed, there are several consequences to the presence of a lapse function with a strong gradient. First of all, the propagation of the waves inside the corotation radius will be slowed down by the 
time dilatation, leading to a longer time to reach saturation. Secondly, the local value of the rotational velocity in that case is actually smaller than it would be for the \refe{Newtonian} case as 
Kerr spacetime leads to a non-vanishing shift vector partially balancing the radial gravitational force. Because of that, we get an actually smaller growth rate as the instability develops 
inside $\sim 4r_g$.
It is interesting to note that those two effects lead to an overall increasing saturation level of the RWI as its develops to smaller and smaller gravitational radii. 

This last point is of particular interest to black hole variability observations. 
Indeed, while we have seen a difference in the time needed to reach saturation, it would require a massive black hole for it to be detectable by nowadays instrument. 
By having some links between the highest possible saturation level, hence ultimately the detectability of the variability, and the spin of the central object, we can place some small
constraints on the spin. In order to do this we would need to create synthetic observations of our simulations in a far away observer frame taking into account general relativistic effects upon the radiative emission of the disc. Such work lies beyond the scope of this paper and will be studied in a forthcoming publication  \citet{Var18}.

\section*{Acknowledgments}
 We acknowledge the financial support from the UnivEarthS Labex program of Sorbonne Paris Cit\'e (ANR-10-LABX-0023 and ANR-11-IDEX-0005-02). The authors would like to thank the referee Jiri Horak for his insightful report that helped improving our paper.

\bibliographystyle{mn2e}

\bibliography{biblio}

\begin{thebibliography}{}
\makeatletter
\relax
\def\mn@urlcharsother{\let\do\@makeother \do\$\do\&\do\#\do\^\do\_\do\%\do\~}
\def\mn@doi{\begingroup\mn@urlcharsother \@ifnextchar [ {\mn@doi@}
  {\mn@doi@[]}}
\def\mn@doi@[#1]#2{\def\@tempa{#1}\ifx\@tempa\@empty \href
  {http://dx.doi.org/#2} {doi:#2}\else \href {http://dx.doi.org/#2} {#1}\fi
  \endgroup}
\def\mn@eprint#1#2{\mn@eprint@#1:#2::\@nil}
\def\mn@eprint@arXiv#1{\href {http://arxiv.org/abs/#1} {{\tt arXiv:#1}}}
\def\mn@eprint@dblp#1{\href {http://dblp.uni-trier.de/rec/bibtex/#1.xml}
  {dblp:#1}}
\def\mn@eprint@#1:#2:#3:#4\@nil{\def\@tempa {#1}\def\@tempb {#2}\def\@tempc
  {#3}\ifx \@tempc \@empty \let \@tempc \@tempb \let \@tempb \@tempa \fi \ifx
  \@tempb \@empty \def\@tempb {arXiv}\fi \@ifundefined
  {mn@eprint@\@tempb}{\@tempb:\@tempc}{\expandafter \expandafter \csname
  mn@eprint@\@tempb\endcsname \expandafter{\@tempc}}}

\bibitem[\protect\citeauthoryear{{Casse}, {Varniere}  \& {Meliani}}{{Casse}
  et~al.}{2017}]{Cass17}
{Casse} F.,  {Varniere} P.,   {Meliani} Z.,  2017, \mn@doi [\mnras]
  {10.1093/mnras/stw2572}, \href
  {http://cdsads.u-strasbg.fr/abs/2017MNRAS.464.3704C} {464, 3704}

\bibitem[\protect\citeauthoryear{{Kerr}}{{Kerr}}{1963}]{Kerr63}
{Kerr} R.~P.,  1963, \mn@doi [Physical Review Letters]
  {10.1103/PhysRevLett.11.237}, \href
  {http://cdsads.u-strasbg.fr/abs/1963PhRvL..11..237K} {11, 237}

\bibitem[\protect\citeauthoryear{{Li}, {Finn}, {Lovelace}  \& {Colgate}}{{Li}
  et~al.}{2000}]{Li00}
{Li} H.,  {Finn} J.~M.,  {Lovelace} R.~V.~E.,   {Colgate} S.~A.,  2000, \mn@doi
  [\apj] {10.1086/308693}, \href
  {http://cdsads.u-strasbg.fr/abs/2000ApJ...533.1023L} {533, 1023}

\bibitem[\protect\citeauthoryear{{Li}, {Colgate}, {Wendroff}  \& {Liska}}{{Li}
  et~al.}{2001}]{Li01}
{Li} H.,  {Colgate} S.~A.,  {Wendroff} B.,   {Liska} R.,  2001, \mn@doi [\apj]
  {10.1086/320241}, \href {http://cdsads.u-strasbg.fr/abs/2001ApJ...551..874L}
  {551, 874}

\bibitem[\protect\citeauthoryear{{Lin}}{{Lin}}{2012}]{Lin12}
{Lin} M.-K.,  2012, \mn@doi [\mnras] {10.1111/j.1365-2966.2012.21955.x}, \href
  {http://cdsads.u-strasbg.fr/abs/2012MNRAS.426.3211L} {426, 3211}

\bibitem[\protect\citeauthoryear{{Lovelace} \& {Hohlfeld}}{{Lovelace} \&
  {Hohlfeld}}{1978}]{Lovelace78}
{Lovelace} R.~V.~E.,  {Hohlfeld} R.~G.,  1978, \mn@doi [\apj] {10.1086/156004},
  \href {http://cdsads.u-strasbg.fr/abs/1978ApJ...221...51L} {221, 51}

\bibitem[\protect\citeauthoryear{{Lovelace} \& {Romanova}}{{Lovelace} \&
  {Romanova}}{2014}]{Lovelace14}
{Lovelace} R.~V.~E.,  {Romanova} M.~M.,  2014, \mn@doi [Fluid Dynamics
  Research] {10.1088/0169-5983/46/4/041401}, \href
  {http://cdsads.u-strasbg.fr/abs/2014FlDyR..46d1401L} {46, 041401}

\bibitem[\protect\citeauthoryear{{Lovelace}, {Li}, {Colgate}  \&
  {Nelson}}{{Lovelace} et~al.}{1999}]{Lovelace99}
{Lovelace} R.~V.~E.,  {Li} H.,  {Colgate} S.~A.,   {Nelson} A.~F.,  1999,
  \mn@doi [\apj] {10.1086/306900}, \href
  {http://cdsads.u-strasbg.fr/abs/1999ApJ...513..805L} {513, 805}

\bibitem[\protect\citeauthoryear{{Lyra} \& {Mac Low}}{{Lyra} \& {Mac
  Low}}{2012}]{Lyra12}
{Lyra} W.,  {Mac Low} M.-M.,  2012, \mn@doi [\apj]
  {10.1088/0004-637X/756/1/62}, \href
  {http://cdsads.u-strasbg.fr/abs/2012ApJ...756...62L} {756, 62}

\bibitem[\protect\citeauthoryear{{Mathews}}{{Mathews}}{1971}]{Mat71}
{Mathews} W.~G.,  1971, \mn@doi [\apj] {10.1086/150883}, \href
  {http://cdsads.u-strasbg.fr/abs/1971ApJ...165..147M} {165, 147}

\bibitem[\protect\citeauthoryear{{Meheut}, {Casse}, {Varniere}  \&
  {Tagger}}{{Meheut} et~al.}{2010}]{Meheut10}
{Meheut} H.,  {Casse} F.,  {Varniere} P.,   {Tagger} M.,  2010, \mn@doi [\aap]
  {10.1051/0004-6361/201014000}, \href
  {http://cdsads.u-strasbg.fr/abs/2010A%26A...516A..31M} {516, A31}

\bibitem[\protect\citeauthoryear{{Meheut}, {Keppens}, {Casse}  \&
  {Benz}}{{Meheut} et~al.}{2012}]{Meheut12}
{Meheut} H.,  {Keppens} R.,  {Casse} F.,   {Benz} W.,  2012, \mn@doi [\aap]
  {10.1051/0004-6361/201118500}, \href
  {http://cdsads.u-strasbg.fr/abs/2012A%26A...542A...9M} {542, A9}

\bibitem[\protect\citeauthoryear{{Meliani}, {Sauty}, {Tsinganos}  \&
  {Vlahakis}}{{Meliani} et~al.}{2004}]{Meliani04}
{Meliani} Z.,  {Sauty} C.,  {Tsinganos} K.,   {Vlahakis} N.,  2004, \mn@doi
  [\aap] {10.1051/0004-6361:20035653}, \href
  {http://cdsads.u-strasbg.fr/abs/2004A%26A...425..773M} {425, 773}

\bibitem[\protect\citeauthoryear{{Mignone} \& {McKinney}}{{Mignone} \&
  {McKinney}}{2007}]{Mignone07}
{Mignone} A.,  {McKinney} J.~C.,  2007, \mn@doi [\mnras]
  {10.1111/j.1365-2966.2007.11849.x}, \href
  {http://cdsads.u-strasbg.fr/abs/2007MNRAS.378.1118M} {378, 1118}

\bibitem[\protect\citeauthoryear{{Misner}, {Thorne}  \& {Wheeler}}{{Misner}
  et~al.}{1973}]{MTW73}
{Misner} C.~W.,  {Thorne} K.~S.,   {Wheeler} J.~A.,  1973, {Gravitation}

\bibitem[\protect\citeauthoryear{{Remillard} \& {McClintock}}{{Remillard} \&
  {McClintock}}{2006}]{Remillard06}
{Remillard} R.~A.,  {McClintock} J.~E.,  2006, \mn@doi [annual review of
  astronomy and astrophysics] {10.1146/annurev.astro.44.051905.092532PDF:
  http://arjournals.annualreviews.org/doi/pdf/10.1146/annurev.astro.44.051905.092532},
  \href
  {http://adsabs.harvard.edu/cgi-bin/nph-bib_query?bibcode=2006ARA%26A..44...49R&db_key=AST}
  {44, 49}

\bibitem[\protect\citeauthoryear{Synge}{Synge}{1957}]{Syn57}
Synge J.,  1957, The relativistic gas.
North-Holland Pub. Co.

\bibitem[\protect\citeauthoryear{{Tagger} \& {Melia}}{{Tagger} \&
  {Melia}}{2006}]{Tagger06}
{Tagger} M.,  {Melia} F.,  2006, \mn@doi [\apjl] {10.1086/499806}, \href
  {http://cdsads.u-strasbg.fr/abs/2006ApJ...636L..33T} {636, L33}

\bibitem[\protect\citeauthoryear{{Tagger} \& {Varni{\`e}re}}{{Tagger} \&
  {Varni{\`e}re}}{2006}]{TV06}
{Tagger} M.,  {Varni{\`e}re} P.,  2006, \mn@doi [\apj] {10.1086/508318}, \href
  {http://cdsads.u-strasbg.fr/abs/2006ApJ...652.1457T} {652, 1457}

\bibitem[\protect\citeauthoryear{{Taub}}{{Taub}}{1948}]{Taub48}
{Taub} A.~H.,  1948, \mn@doi [Physical Review] {10.1103/PhysRev.74.328}, \href
  {http://cdsads.u-strasbg.fr/abs/1948PhRv...74..328T} {74, 328}

\bibitem[\protect\citeauthoryear{{Varni{\`e}re} \& {Tagger}}{{Varni{\`e}re} \&
  {Tagger}}{2006}]{VT06}
{Varni{\`e}re} P.,  {Tagger} M.,  2006, \mn@doi [\aap]
  {10.1051/0004-6361:200500226}, \href
  {http://cdsads.u-strasbg.fr/abs/2006A%26A...446L..13V} {446, L13}

\bibitem[\protect\citeauthoryear{{Varniere}, {Tagger}  \&
  {Rodriguez}}{{Varniere} et~al.}{2011}]{VTR11}
{Varniere} P.,  {Tagger} M.,   {Rodriguez} J.,  2011, \mn@doi [\aap]
  {10.1051/0004-6361/201015028}, \href
  {http://cdsads.u-strasbg.fr/abs/2011A%26A...525A..87V} {525, A87}

\bibitem[\protect\citeauthoryear{{Varni{\`e}re}, {Tagger}  \&
  {Rodriguez}}{{Varni{\`e}re} et~al.}{2012}]{VTR12}
{Varni{\`e}re} P.,  {Tagger} M.,   {Rodriguez} J.,  2012, \mn@doi [\aap]
  {10.1051/0004-6361/201116698}, \href
  {http://cdsads.u-strasbg.fr/abs/2012A%26A...545A..40V} {545, A40}

\bibitem[\protect\citeauthoryear{{Varniere}, {Casse}  \& {Vincent}}{{Varniere}
  et~al.}{2018}]{Var18}
{Varniere} P.,  {Casse} F.,   {Vincent} F.,  2018, in prep.

\bibitem[\protect\citeauthoryear{{Vincent}, {Meheut}, {Varniere}  \&
  {Paumard}}{{Vincent} et~al.}{2013}]{Vin13}
{Vincent} F.~H.,  {Meheut} H.,  {Varniere} P.,   {Paumard} T.,  2013, \mn@doi
  [\aap] {10.1051/0004-6361/201220695}, \href
  {http://cdsads.u-strasbg.fr/abs/2013A%26A...551A..54V} {551, A54}

\bibitem[\protect\citeauthoryear{{Vincent}, {Paumard}, {Perrin}, {Varniere},
  {Casse}, {Eisenhauer}, {Gillessen}  \& {Armitage}}{{Vincent}
  et~al.}{2014}]{Vin14}
{Vincent} F.~H.,  {Paumard} T.,  {Perrin} G.,  {Varniere} P.,  {Casse} F.,
  {Eisenhauer} F.,  {Gillessen} S.,   {Armitage} P.~J.,  2014, \mn@doi [\mnras]
  {10.1093/mnras/stu812}, \href
  {http://cdsads.u-strasbg.fr/abs/2014MNRAS.441.3477V} {441, 3477}

\makeatother
\end{thebibliography}

\end{document}